\documentclass[aps,prb,onecolumn,amsmath,amssymb,superscriptaddress,floatfix,eqsecnum]{revtex4-2}

\usepackage{float}
\usepackage{physics}
\usepackage{times}
\usepackage{bm}
\usepackage{textcomp}
\usepackage{graphicx}
\allowdisplaybreaks 
\usepackage{braket}
\usepackage{array}
\usepackage[breaklinks]{hyperref}
\usepackage{subdepth}
\usepackage{mathtools}
\usepackage{cancel,soul,ulem}
\usepackage{wasysym}

\hypersetup{colorlinks=true, linkcolor=blue, citecolor=blue, filecolor=blue, urlcolor=blue}

\newcommand{\pd}{\phantom{\dag}}

\newcommand{\RN}[1]{\textup{\uppercase\expandafter{\romannumeral#1}}}

\begin{document}

\title{A practical method to detect, analyse and engineer higher order Van Hove singularities in multi-band Hamiltonians}

\author{Anirudh Chandrasekaran}
\affiliation{Department of Physics and Centre for the Science of Materials,
Loughborough University,  Loughborough LE11 3TU, UK.}

\author{Joseph J. Betouras}
\affiliation{Department of Physics and Centre for the Science of Materials,
Loughborough University,  Loughborough LE11 3TU, UK.}




\date{\rm\today}

\begin{abstract}
We present a practical method to detect, diagnose and engineer higher order Van Hove singularities in multiband systems, with no restrictions on the number of bands and hopping terms. The method allows us to directly compute the Taylor expansion of the dispersion of any band at arbitrary points in momentum space, using a generalised extension of the Feynman Hellmann theorem, which we state and prove. Being fairly general in scope, it also allows us to incorporate and analyse the effect of tuning parameters on the low energy dispersions, which can greatly aid the engineering of higher order Van Hove singularities. A certain class of degenerate bands can be handled within this framework. We demonstrate the use of the method, by applying it to the Haldane model. 

\end{abstract}

\maketitle


\section{Introduction}
Nearly a century after its introduction, modelling of the band structure of crystalline solids continues to play a significant role in our understanding of real materials. Much progress has been made, both in the realm of techniques for the calculation of band structures, and the analysis of non-trivial band features. Progress in the latter is particularly epitomized by the rapid advances made in recent times, in the areas of band topology and geometry. The investigation of unusual band topologies and their physical importance was initiated by Lifshitz and Van Hove \cite{Lifshitz,vanHove}. Their work led to the notions of Fermi surface topological transitions, where the Fermi surface geometry changes suddenly when some parameters in the system are tuned and the associated density of states (DOS) divergences. In his pioneering early work, Lishitz identified two types of Fermi surface topological transitions, namely the pocket appearing/disappearing type and the neck formation/collapse type~\cite{Lifshitz}. Usually in Fermi surface topological transitions, the Fermi surface hosts one or more critical points of the dispersion.

Critical points are accompanied by a vanishing gradient of the energy dispersion, and typically correspond to either a maximum, a minimum or a saddle point of the dispersion. For a two dimensional system, the canonical form of the Taylor expansion of the dispersion at a critical point to quadratic order, is normally given by $\pm k_x^2 \pm k_y^2$. Such a form of the dispersion leads to a logarithmic divergence in the DOS~\cite{vanHove} at the critical energy. Sometimes however, we need to expand the dispersion beyond quadratic order to adequately capture the low energy features. These are situations where the determinant of the Hessian matrix of the dispersion, evaluated at the critical point, also vanishes. In such a scenario, the critical point is referred to as a higher order critical point, and the dispersion is said to have a higher order Van Hove singularity (HOVHS). A simple and commonly reported example of HOVHS is the cusp singularity with dispersion $k_x^4 - k_y^{2}$~\cite{Chandrasekaran-Shtyk-Betouras-Chamon, Efremov-Betouras, Shtyk}. 

In the case of HOVHS, the divergence of the DOS is enhanced to a much stronger power law, often with asymmetric prefactors above and below the critical energy.~\cite{Chandrasekaran-Shtyk-Betouras-Chamon, LiangFu}. This is expected to strengthen electronic correlations in the vicinity of the HOVHS. The consequences of this could be potentially dramatic \cite{Zervou-Goldstein-Efremov-Betouras}, and may lead to novel quantum phases. The fundamental reason is that due to the vanishing gradient of the energy dispersion in the vicinity of those points, the fermions get heavy and interaction effects become important.

Ordinary Van Hove singularities have been observed in the context of Lifshitz transitions in many materials including cuprates and iron based superconductors, cobaltates, $\text{Sr}_{2} \text{RuO}_{4}$ and heavy fermion systems~\cite{Aoki,Barber,Bernhabib,Khan,Coldea,Okamoto,Sherkunov-Chubukov-Betouras,Slizovskiy-Chubukov-Betouras,Stewart,Yelland}. With the recent advances in the experimental methods of tuning materials and improved resolutions of various spectroscopic probes (particularly angle resolved photoemission spectroscopy (ARPES)), there is both an increasing interest and emerging avenues for the investigation of the presence and consequence of exotic band geometry in real materials. This is particularly true for HOVHS, which have been linked with the unusual magnetic and thermodynamic behaviour of $\mathrm{Sr_3 Ru_2 O_7}$~\cite{Efremov-Betouras}, twisted bilayer graphene near magic angle, close to half filling~\cite{Yuan, Sherkunov-Betouras, Classen_2020}, the recently proposed unusual quantum phase referred to as `supermetal', which has diverging susceptibilities without the presence of any long range order~\cite{Isobe}, the distinct Landau level spectrum of biased bilayer graphene \cite{Shtyk} and the `extended' Van Hove singularity in both doped graphene~\cite{doped_graphene} and highly overdoped graphene~\cite{Rosenzweig_2020}.  More recently, HOVHS have been reported in the kagome superconductors\cite{Kang, kagome2, Consiglio, Neupert}, and may be relevant to the observed phases of doped Bernal bilayer graphene ~\cite{Zhou_2021}.

It is important to note that while a regular VHS is simply an ordinary saddle point singularity (having canonical dispersion $k_x^2 - k_y^2$), HOVHS are, in contrast, a large class of \emph{distinct} singularities. Although in principle there are infinitely many of them, we need to consider only a finite subset of singularities when studying two dimensional and layered materials (see Ref~\cite{Chandrasekaran-Shtyk-Betouras-Chamon} for a justification, and Refs~\cite{Chandrasekaran-Shtyk-Betouras-Chamon, LiangFu} for a thorough classification of these relevant singularities). Apart from this, another complication that arises in the context of HOVHS is that they are more elusive compared to regular VHS, i.e, they typically require a fine tuning of parameters in the system (by the application of strain, pressure, twist, bias voltage, doping, etc). In addition, HOVHS may be sensitive to impurities and changes in the parameters, upon which the higher order critical point could split into a number of ordinary critical points. As a recent study of the effect of impurities on HOVHS~\cite{Chandrasekaran-Betouras} shows, for a low concentration of impurities, the distinct quantitative signatures of the HOVHS do survive. 

As already noted, there has been a renewed interest and improved ability to experimentally tune real materials by various probes. This renders a sound theoretical analysis of HOVHS a highly relevant and necessary task. To this end, in this work, we address the question of how we can detect a HOVHS given a k-space Hamiltonian, even when the description is detailed (in the sense of number of terms, hopping parameters, etc.) and involves multiple bands. The HOVHS can be diagnosed using the Taylor expansion of the band in question, at the chosen $\vb{k}$ point.  Low energy theories determined either by polynomial fits or purely symmetry considerations are often also a stepping stone for further analysis, such as diagrammatic many-body treatment of electronic correlations and transport studies \cite{Zervou-Goldstein-Efremov-Betouras, Kokkinis, Chandrasekaran-Betouras}. But the full and correct Taylor expansion to a given order is needed for the unambiguous diagnosis of a HOVHS (see Ref~\cite{Chandrasekaran-Shtyk-Betouras-Chamon} for an illustration of how relying purely on symmetry considerations can lead to a misdiagnosis of the HOVHS, the DOS divergence exponent and the ratio of prefactors). 


While HOVHS have been increasingly reported in real materials and theoretical models, detecting their presence and diagnosing their precise nature is not a straightforward task. Even when the Taylor expansion of the band at the critical point is available, naive inspection or even scaling arguments may not help us identify the correct singularity, and its associated power law exponents and ratio of prefactors (see Ref~\cite{Chandrasekaran-Shtyk-Betouras-Chamon} for a discussion of the issue with an illustrative example). Nevertheless, a straightforward algorithm exists, which, given the Taylor expansion, can be used to exactly pinpoint the singularity that is present~\cite{Chandrasekaran-Shtyk-Betouras-Chamon}. To this end it becomes necessary for us to compute the Taylor expansion of the bands in tight-binding and Wannierised models, particularly when these models are able to account for experiments on materials (such as angle resolved photoemission spectroscopy and quasiparticle interference measurements). 

What is widely used in this context is polynomial fits to the band along some directions in k-space. While this may indeed reveal some aspects of the Taylor expansion, it may at times altogether miss the presence of a higher order singularity if the directions of the fits do not coincide with the `principal' directions of the singularity. For example, the singularity with dispersion relation: 
\begin{equation}
\varepsilon({\bf k}) \propto (k_x^{\,} + k_y^{\,})^4 - (k_x^{\,} - k_y^{\,})^2
\end{equation}
will not be revealed as a higher order saddle by polynomial fits of the dispersion along the $k_x^{\,}$ and $k_y^{\,}$ directions. Even in the cases where the polynomial fit reveals a higher order singularity, by virtue of absence or smallness of the quadratic part of the fit, the full Taylor expansion to a specified order may not be revealed by polynomial fits along some chosen directions. In fact, the fitted polynomial may even have very different coefficients compared to the Taylor expansion arising from a tight-binding model that adequately describes the same dispersion. Therefore, given a tight-binding model, we need a general technique for extracting the Taylor expansion of bands at arbitrary k-points unambiguously. 

In this paper, we provide a technique based on an extension of the Feynman Hellmann theorem, to directly compute the Taylor expansion to any order, of band dispersion given the k-space Hamiltonian, with no restriction on number of bands and hopping terms. The method can be used to treat a case of degenerate bands, although it can be extended to any case. An explicit proof of the method, to any degree of Taylor expansion is provided in the appendix. 
We first deal with multi-band models in section~\ref{sec:tight_binding} where we provide the method and some general formulae. It is worth emphasizing that the method can be applied to any Hamiltonian $H(\vb{k})$, not just ones arising from tight binding models. In Sec.~\ref{sec:haldane} we apply the method and the generic results to the case of the Haldane model tuned to a higher order singularity. Finally, we conclude in Sec.~\ref{sec:conclusion}. For clarity and to facilitate the flow, details of the calculations and the full general proof are presented in the appendix. 

\section{Multi-band tight binding models}
\label{sec:tight_binding}

As described in the introduction, the active bands near the Fermi level of many real materials can be captured by complicated tight binding and $k \cdot p$ models that might, in practice, require a large number of Wannier functions or orbitals and several hopping terms to be able to adequately reproduce DFT and ARPES bands. This can often make the analysis of low energy theory around some ${\bf k}$ point quite difficult, especially if we want to avoid resorting to polynomial fits or if we are interested in tracking the effect of changing tuning parameters on the dispersion. Since the characteristic equation has a degree equal to the dimension of the Hamiltonian matrix (i.e the total number of bands), any analytic solution to the eigenvalue problem for general $\vb{k}$, is possible only up to the four band case, since the general quintic polynomial (of degree $5$) and higher order polynomials can not be solved in terms of radicals~\cite{abel}. Already for cubic and quartic polynomials, the general solutions are quite complicated. We therefore need a more practical method to obtain the Taylor expansion around a particular $\vb{k}$ point in a specified band. In the next section we describe such a method that can handle large and detailed k-space Hamiltonians (including those based on TBMs) and can be readily implemented numerically.

\subsection{Preliminaries}
To compute the Taylor expansion to order $N$ at $\vb{k}_0^{\,}$, in principle we need to compute all the mixed partial derivatives $\partial_1^{l_1} \partial_2^{l_2} \cdots \partial_d^{l_d} \varepsilon_n (\vb{k})|_{\vb{k}_0}$ for $1 \leqslant l_1^{\,} + l_2^{\,} + \cdots + l_d^{\,} \leqslant N$ in a $d$-dimensional system. However, computing all these mixed partial derivatives within the Feynman Hellmann method will be rather involved. We can simplify this procedure considerably by computing instead, the derivatives with respect to a \emph{single} variable $\lambda$, taking the form $\partial^M \varepsilon_n^{\,} (\vb{k}_0^{\,} + \lambda \, \vb{k}) / \partial \lambda^M $ and setting $\lambda$ to zero at the end. By repeated application of the chain rule for differentiation, it can be shown that
\begin{equation}
\frac{\partial^M \varepsilon_n^{\,} (\vb{k}_0^{\,} + \lambda \, \vb{k})}{\partial \lambda^M} \bigg|_{\lambda = 0} = \sum_{\substack{1 \leqslant l_1, l_2, \cdots , l_d < M \\ l_1 + l_2 + \cdots + l_d = M}} \frac{M!}{l_1^{\,}! \, l_2^{\,}! \cdots l_d^{\,}!}
\,
\frac{\partial^M \varepsilon_n^{\,} (\vb{k})}{\partial k_1^{l_1} \, \partial k_2^{l_2} \cdots \partial k_d^{l_d}} \bigg|_{\vb{k} = \vb{k}_0} \, k_1^{l_1} \, k_2^{l_2} \cdots k_d^{l_d}.
\end{equation}
Thus, $\frac{1}{M!} \frac{\partial^M}{\partial \lambda^M} \varepsilon_n^{\,} (\vb{k}_0^{\,} + \lambda \, \vb{k}) \bigg|_{\lambda = 0}$ is the sum of \emph{all} the monomials of degree $M$ in the Taylor expansion of $\varepsilon_n^{\,} (\vb{k})$ at $\vb{k}_0^{\,}$. The full Taylor expansion to order $N$ is then given by
\begin{equation}
\varepsilon_n^{\,} (\vb{k}_0^{\,} + \vb{k}) = \sum_{M = 1}^{N} \frac{1}{M!} \, \frac{\partial^M \varepsilon_n^{\,} (\vb{k}_0^{\,} + \lambda \, \vb{k})}{\partial \lambda^M} \bigg|_{\lambda = 0} + \mathcal{O} \left( k^{N+1} \right).
\label{eq:tayl_expansion_1}
\end{equation}
In order to compute the Taylor expansion of the band dispersion, it is sufficient to use this trick along with the Feynman Hellmann formula, whose extension to higher orders will be stated and derived below.


\subsection{The method}\label{sec:method}

Before proceeding to the method, we state the important assumptions. The Hamiltonian is a finite dimensional matrix that is a smooth function of $\vb{k}$. This is always the case with tight-binding models, however large and complicated they may be. We allow for the band in question to be degenerate with some other bands but demand that its Taylor expansion to order $N$ also coincides with the other bands. This is a restrictive assumption which will be relaxed in a future work when the method will be extended to treat the more complicated scenarios of bands touching and intersecting each other in complicated ways (like a linear band crossing for example. Such multi-band intersections, sometimes collectively referred to as multi-fold fermions, can involve several bands and may be stabilized by symmetries.~\cite{bradlyn2016beyond}).

We are interested in computing the Taylor expansion at $\vb{k}_0^{\,}$ to order $N$, of the $n^{\text{th}}$ band (denoted by $\varepsilon_n^{\,}(\vb{k})$) arising from the Hamiltonian matrix $\widehat{H}(\vb{k})$. The method proceeds to compute the terms order by order.
\begin{enumerate}
\item Compute numerically, all the eigenvalues and eigenvectors of $\widehat{H}(\vb{k})$ at the chosen point $\vb{k}_0^{\,}$. Let us order these eigenvalues and denote them by $\varepsilon_1^{\,}$, $\varepsilon_2^{\,}$,..., $\varepsilon_{\mathcal{N}}^{\,}$. The corresponding eigenvectors are arranged into the columns of a matrix $\widehat{E}$.

\item Perform a change of basis from the orbital basis to the eigenbasis \emph{at} $\vb{k}_0^{\,}$ by transforming the Hamiltonian: $\widehat{H}(\vb{k}) \rightarrow \widehat{E}^{\dag} \, \widehat{H}(\vb{k}) \, \widehat{E}$. (Note that we are transforming the Hamiltonian as a function of $\vb{k}$, not just the Hamiltonian at the chosen $\vb{k}_0^{\,}$. The eigenvectors that constitute the new basis are however the numerically computed ones at $\vb{k}_0^{\,}$. Technically this change of basis is not needed. But it makes the algorithm more efficient since we will not have to repeatedly compute matrix elements of the form $\langle m_1^{\,} | \partial^{M} \widehat{H} | m_2^{\,} \rangle$ with respect to the eigenvectors $|m_1^{\,} \rangle$ and $|m_2^{\,} \rangle$ by matrix multiplication, and can instead directly read off the corresponding matrix elements).

\item Calculate the following matrices analytically by directly differentiating the basis transformed Hamiltonian:
\begin{equation}
\partial \widehat{H} \coloneqq \frac{\partial \widehat{H}(\vb{k}_0^{\,} + \lambda \, \vb{k})}{\partial \lambda} \bigg|_{\lambda = 0}, \partial^2 \widehat{H} \coloneqq \frac{\partial^2 \widehat{H}(\vb{k}_0^{\,} + \lambda \, \vb{k})}{\partial \lambda^2} \bigg|_{\lambda = 0}, \cdots , \partial^N \widehat{H} \coloneqq \frac{\partial^N \widehat{H}(\vb{k}_0^{\,} + \lambda \vb{k})}{\partial \lambda^N} \bigg|_{\lambda = 0} .
\label{eq:Hamltn_derivs}
\end{equation}
These matrices when divided by the appropriate factorial weights and assembled together make up the Taylor expansion of the Hamiltonian at $\vb{k}_0^{\,}$ to order $N$.

\item To compute the Taylor expansion, we shall now outline the procedure to compute the following polynomials and assemble them into the Taylor series as in Eq~\ref{eq:tayl_expansion_1}:
\begin{equation}
\partial \varepsilon_n^{\,} \coloneqq \frac{\partial \varepsilon_n^{\,}(\vb{k}_0^{\,} + \lambda \, \vb{k})}{\partial \lambda} \bigg|_{\lambda = 0}, \partial^2 \varepsilon_n^{\,} \coloneqq \frac{\partial^2 \varepsilon_n^{\,}(\vb{k}_0^{\,} + \lambda \, \vb{k})}{\partial \lambda^2} \bigg|_{\lambda = 0}, \cdots , \partial^N \varepsilon_n^{\,} \coloneqq \frac{\partial^N \varepsilon_n^{\,}(\vb{k}_0^{\,} + \lambda \, \vb{k})}{\partial \lambda^N} \bigg|_{\lambda = 0} .
\end{equation}

\item Before computing each of the above polynomials, we formally define the following recursive function for $1 \leqslant m, n \leqslant \dim \widehat{H}$, with $\varepsilon_m^{\,} \neq \varepsilon_n^{\,}$ and $\kappa \geqslant 1$:
\begin{equation}
f(\kappa, m, n) = 
 \frac{\left[ \partial^{\kappa} \widehat{H} \right]_{m,n}^{\,}}{\varepsilon_n^{\,} - \varepsilon_m^{\,}} 
 + \sum\limits_{\widetilde{\kappa} = 1}^{\kappa - 1} \begin{pmatrix}
\kappa \\
\widetilde{\kappa}
\end{pmatrix} 
\sum\limits_{| \widetilde{m} \rangle \notin D(n)} 
\frac{
\left[ \partial^{\kappa-\widetilde{\kappa}} \widehat{H} \right]_{m,\widetilde{m}}^{\,} - \left( \partial^{\kappa-\widetilde{\kappa}}\varepsilon_n^{\,} \right) \, \delta_{m,\widetilde{m}}^{\,}
}{\varepsilon_n^{\,} - \varepsilon_m^{\,}} \, f \left( \widetilde{\kappa}, \widetilde{m}, n \right),
\label{eq:recursive_func}
\end{equation}
where $[\partial^{\kappa} \widehat{H}]_{m,n}^{\,}$ is a shorthand for $\langle m | \partial^{\kappa} \widehat{H} |n \rangle$, a matrix element of the matrix $\partial^{\kappa} \widehat{H}$,  
$\begin{pmatrix}
\kappa \\
\widetilde{\kappa}
\end{pmatrix} $ is the binomial coefficient and $D(n)$ is the set of all the eigenstates that are degenerate to $|n\rangle$ (including itself), so that the $|\widetilde{m}\rangle$ sum runs over all the eigenvectors at $\vb{k}_0^{\,}$ that are \emph{not} degenerate to $|n\rangle$.

Given that this is a recursive definition, we will obtain a sequence of product of the function $f$ evaluated at different $\widetilde{m}$ and decreasing $\widetilde{\kappa}$ in the full expansion of the right hand side. We list the explicit expression for $f(\kappa,m,n)$ for $\kappa = 1, 2$
\begin{subequations}
\begin{align}
f(1,m,n) =& \frac{[\partial \widehat{H}]_{m,n}}{\varepsilon_n^{\,} - \varepsilon_m^{\,}}, \\ 
f(2,m,n) =& \frac{[\partial^2 \widehat{H}]_{m,n}}{\varepsilon_n^{\,} - \varepsilon_m^{\,}} + 2 \sum_{|\widetilde{m}\rangle \notin D(n)} \frac{[\partial \widehat{H}]_{n,\widetilde{m}}}{\varepsilon_n^{\,} - \varepsilon_m^{\,}} \, \frac{[\partial \widehat{H}]_{\widetilde{m},n}}{\varepsilon_n^{\,} - \varepsilon_{\widetilde{m}}^{\,}} - 2 \, \frac{\partial \varepsilon_n^{\,} \, [\partial \widehat{H}]_{m,n}}{\left( \varepsilon_n^{\,} - \varepsilon_{m}^{\,} \right)^2}.
\end{align}
\end{subequations}

The sequence of product of $f$'s always terminates when $\widetilde{\kappa}=1$, and we have in that case $f \left( \widetilde{\kappa} = 1, \widetilde{m}, n \right) = [\partial \widehat{H}]_{\widetilde{m},n}^{\,} / (\varepsilon_n^{\,} - \varepsilon_{\widetilde{m}}^{\,})$. For any $\kappa > 1$ and given $m$ and $n$, to evaluate $f(\kappa, m, n)$, we need to know all the lower derivatives of the dispersion of band $n$, denoted by $\partial \varepsilon_n^{\,}, \partial^2 \varepsilon_n^{\,}, \cdots , \partial^{\kappa-1} \varepsilon_n^{\,}$. But since we know the first derivative from Feynman - Hellmann lemma, $\partial \varepsilon_n^{\,} = \left[ \partial \widehat{H} \right]_{n,n}^{\,}$, we can explicitly evaluate all the higher derivatives from $\partial^2 \varepsilon_n^{\,}$ onwards order by order using the extended Feynman-Hellmann formula (see Appendix for proof):
\begin{equation}
\partial^M \varepsilon_n^{\,} = \left[ \partial^M \widehat{H} \right]_{n,n}^{\,} + \sum_{\kappa = 1}^{M - 1} \begin{pmatrix}
M \\
\kappa
\end{pmatrix}
\sum_{|m\rangle \notin D(n)} \left[ \partial^{M-\kappa} \widehat{H} \right]_{n,m} \, f(\kappa, m, n).
\label{eq:band_derivative_algorithm}
\end{equation}

\item Having computed $\partial \varepsilon_n^{\,}$ through $\partial^N \varepsilon_n^{\,}$, we assemble them into the Taylor polynomial of degree $N$ at $\vb{k}_0^{\,}$:
\begin{equation}
j^N (\varepsilon_n^{\,}) |_{\vb{k}_0} = \varepsilon_n^{\,} + \frac{\partial \varepsilon_n^{\,}}{1!} + \frac{\partial^2 \varepsilon_n^{\,}}{2!} + \cdots + \frac{\partial^N \varepsilon_n^{\,}}{N!}.
\label{eq:Tayl_expansion_2}
\end{equation}
\end{enumerate}

Before we proceed further, it is important to emphasize that the method can be also used to track the changes in the dispersion due to small changes in the {\it tuning parameters}. This can be achieved by simply redefining $\partial^N \widehat{H}$ in Eq~\ref{eq:Hamltn_derivs}. For example, consider the case where we have the Hamiltonian of the system parametrized for changes in some hopping term $t$ around a critical value $t_0^{\,}$ (this could be the result of twist, strain, pressure etc). To obtain the effect of the tuning directly in the Taylor expansion of the dispersion, we redefine Eq~\ref{eq:Hamltn_derivs} as follows and apply the rest of the algorithm
\begin{multline}
\partial \widehat{H} \coloneqq \frac{\partial \widehat{H}(\vb{k}_0^{\,} + \lambda \, \vb{k}, t_0^{\,} + \lambda \, \delta t)}{\partial \lambda} \bigg|_{\lambda = 0}, \partial^2 \widehat{H} \coloneqq \frac{\partial^2 \widehat{H}(\vb{k}_0^{\,} + \lambda \, \vb{k}, t_0^{\,} + \lambda \, \delta t)}{\partial \lambda^2} \bigg|_{\lambda = 0}, \cdots  \\
\cdots , \partial^N \widehat{H} \coloneqq \frac{\partial^N \widehat{H}(\vb{k}_0^{\,} + \lambda \vb{k}, t_0^{\,} + \lambda \, \delta t)}{\partial \lambda^N} \bigg|_{\lambda = 0} .
\label{eq:Hamltn_derivs_2}
\end{multline}
This will then yield terms in the Taylor expansion that contain $\delta t, \delta t^2, \cdots$ which will track the changes in the dispersion under the tuning of parameters. This procedure is, by no means, restricted to a single tuning parameter. If we search for unusual band geometries under tuning, this procedure can be very helpful. All this is illustrated using an example in section~\ref{sec:haldane}.
 
\subsection{Unpacking the formula}
Before proceeding with the application, we need to state the explicit form of the linear, quadratic and cubic derivatives arising from Eqs~\ref{eq:recursive_func} and~\ref{eq:band_derivative_algorithm} which will be used later. The detailed derivations are included in the Appendix.
\begin{subequations}
\begin{align}
\partial \varepsilon_n^{\,} 
=& \langle n | \partial \widehat{H} | n \rangle , \label{eq:Taylor_first}
\\
\partial^2 \varepsilon_n^{\,} 
=& \langle n | \partial^2 \widehat{H} | n \rangle + 2 \sum_{|m_1 \rangle \notin D(n)} \frac{\langle n | \partial \widehat{H} | m_1^{\,} \rangle \langle m_1^{\,} | \partial \widehat{H} | n \rangle}{\varepsilon_n^{\,} - \varepsilon_{m_1}^{\,}},
\\
\partial^3 \varepsilon_n^{\,} 
=& \langle n | \partial^3 \widehat{H} | n \rangle + 3 \sum_{|m_1 \rangle \notin D(n)} \frac{\langle n | \partial^2 \widehat{H} | m_1^{\,} \rangle \langle m_1^{\,} | \partial \widehat{H} | n \rangle}{\varepsilon_n^{\,} - \varepsilon_{m_1}^{\,}} + 3 \sum_{|m_1 \rangle \notin D(n)} \frac{\langle n | \partial \widehat{H} | m_1^{\,} \rangle \langle m_1^{\,} | \partial^2 \widehat{H} | n \rangle}{\varepsilon_n^{\,} - \varepsilon_{m_1}^{\,}} \nonumber
\\
& \quad + 6 \sum_{|m_1 \rangle \notin D(n)} \sum_{|m_2 \rangle \notin D(n)} \frac{\langle n | \partial \widehat{H} | m_1^{\,} \rangle \langle m_1^{\,} | \partial \widehat{H} | m_2^{\,} \rangle \langle m_2^{\,} | \partial \widehat{H} | n \rangle}{(\varepsilon_n^{\,} - \varepsilon_{m_1}^{\,})(\varepsilon_n^{\,} - \varepsilon_{m_2}^{\,})} \nonumber 
\\
& \quad -6 \, \partial \varepsilon_n^{\,} \sum_{|m_1 \rangle \notin D(n)} \frac{\langle n | \partial \widehat{H} | m_1^{\,} \rangle \langle m_1^{\,} | \partial \widehat{H} | n \rangle}{(\varepsilon_n^{\,} - \varepsilon_{m_1}^{\,})^2}.
\end{align}
\label{eq:Taylor_explicit}
\end{subequations}
Note that the formulae listed above only require us to be able to numerically diagonalize the Hamiltonian $\widehat{H}(\vb{k})$ at a chosen $\vb{k}_0^{\,}$, and be able to differentiate $\widehat{H}(\vb{k})$ analytically a finite number of times, all of which can be readily and quickly done using a computer algebra system (such as Mathematica~\cite{Mathematica} or Maple~\cite{Maple}). Knowing the eigenvalues at $\vb{k}_0^{\,}$ and the matrix elements of various derivatives of $\widehat{H}(\vb{k})$, we shall be able to compute the Taylor expansion order by order, with each higher order depending upon the lower order terms. The first order term (Eq~\ref{eq:Taylor_first}) is just the familiar Feynman - Hellmann result.

\section{Haldane model as an example}\label{sec:haldane}
In this section, we benchmark the method against the widely studied Haldane model~\cite{HaldaneModel}. An extension of the Haldane model can be tuned to host the monkey saddle singularity having dispersion $k_y^3 - 3 \, k_x^2 k_y$~\cite{Chandrasekaran-Betouras}. Although the method can more generally be applied to larger systems that require numerical diagonalization, we choose the Haldane model for its simplicity and analytic tractability, so that it serves as a pedagogical example of how to apply the method. The underlying lattice for the Haldane model is a honeycomb lattice similar to graphene, with two triangular sublattices A and B. The nearest neighbor unit vectors originating from the A sublattice on to the B sublattice are given by
\begin{equation}
\vb{a}_1^{\pd} = \begin{pmatrix} 1 \\ \\ 0 \end{pmatrix}, \vb{a}_2^{\pd} = \begin{pmatrix} -\frac{1}{2} \\ \\ \frac{\sqrt{3}}{2} \end{pmatrix} , \vb{a}_3^{\pd} = \begin{pmatrix} -\frac{1}{2} \\ \\ -\frac{\sqrt{3}}{2} \end{pmatrix}.
\end{equation}

\begin{figure}[t]
\centering
\includegraphics[width=\linewidth]{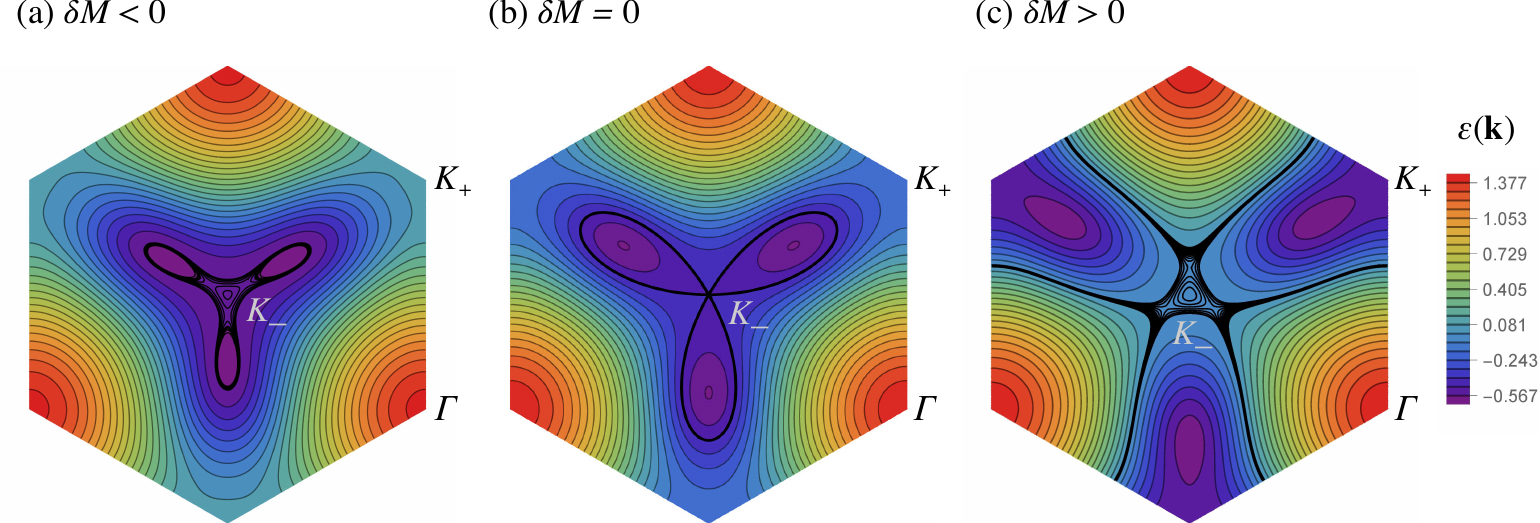}
\caption{The constant energy contours of the Haldane model near critical tuning, in a Brillouin zone centred at the $K_{-}^{\,}$ point. In (a), the staggered chemical potential is tuned slightly below the critical tuning $-M_0^{\,}$. This causes the higher order critical point to split into a number of ordinary critical points: a central minimum surrounded by three regular saddle points. At the critical tuning (i.e when $\delta M = 0$ or equivalently $M = - M_0^{\,}$), we get a three-fold rotation symmetric higher order singularity, known as the monkey saddle. As the staggered chemical potential $M$ is tuned slightly above $-M_0^{\,}$ once again the higher order critical point is split into a number of ordinary critical points: a central maximum surrounded by three regular saddle points. This is depicted in (c).}
\label{fig:Monkey}
\end{figure}

The next nearest neighbor vectors are given by $\vb{b}_1^{\pd} = \vb{a}_2^{\pd} - \vb{a}_3^{\pd}$, $\vb{b}_2^{\pd} = \vb{a}_3^{\pd} - \vb{a}_1^{\pd}$ and $\vb{b}_3^{\pd} = \vb{a}_1^{\pd} - \vb{a}_2^{\pd}$. The full $\vb{k}$-space Hamiltonian is then the sum of nearest neighbor, staggered chemical potential and next nearest neighbor terms: $\widehat{H}(\vb{k}) = \widehat{H}_0^{\pd} (\vb{k}) + M \sigma_z^{\pd} + 2 t_2^{\pd} \sum_i \sin (\vb{k} \cdot \vb{b}_i^{\pd}) \, \sigma_z^{\pd}$, where $\sigma_z^{\pd}$ is the familiar Pauli matrix and
\begin{equation}
\widehat{H}_0^{\pd} (\vb{k}) = \begin{pmatrix}
0 & t_1^{\pd} \sum_i e^{i \vectorbold{k} \cdot \vectorbold{a}_i} \\
t_1^{\pd} \sum_i e^{-i \vectorbold{k} \cdot \vectorbold{a}_i} & 0
\end{pmatrix}.
\end{equation}
Since the Hamiltonian is two-dimensional, it can be diagonalized exactly to give two bands indexed by $n = 1, 2$
\begin{equation}
\varepsilon_{n}^{\pd} (\vectorbold{k}) = (-1)^n \bigg[ \bigg(M + 2 t_2 \sum_i \sin (\vectorbold{k} \cdot \vectorbold{b}_i^{\pd}) \bigg)^2 + t_1^2 \bigg| \sum_i e^{i \vectorbold{k} \cdot \vectorbold{a}_i^{\pd}} \bigg|^2 \bigg]^{\frac{1}{2}} .
\end{equation}
Consider the following point of three-fold rotation symmetry
\begin{equation}
\vb{K}_-^{\pd} = \frac{4\pi}{3\sqrt{3}} \begin{pmatrix}
0 \\ \\
1
\end{pmatrix}.
\end{equation}
Let us define a critical value for the staggered chemical potential
\begin{equation}
M_0^{\,} = \frac{t_1^2 - 18 \, t_2^2}{2 \, \sqrt{3} \, t_2^{\pd}}.
\end{equation}
We can analytically derive the Taylor expansion for the upper band $(n=2)$ dispersion, which around the $K_-^{\pd}$ point reads
\begin{align}
\varepsilon_2^{\pd} (\vb{k} + \vb{K}_-^{\pd}) \approx & \left| M - 3\sqrt{3} \, t_2^{\,} \right| + \frac{9 \sqrt{3} \, t_2^{\,} (M + M_0^{\,})}{4 \left| M - 3\sqrt{3} \, t_2^{\,} \right|} \left( k_x^2 + k_y^2 \right) \nonumber \\
& \quad + \frac{6 \, t_2^{\,} \left( \sqrt{3} \, M - 9 \, t_2^{\,} \right) - 9 \, t_1^2}{16 \left| M - 3\sqrt{3} \, t_2^{\,} \right|} \left( k_y^3 - 3 \, k_y^{\,} \, k_x^2 \right) + \mathcal{O} \left( k^4 \right).
\end{align}
When we tune $M \rightarrow - M_0^{\,}$
\begin{align}
\varepsilon_2^{\pd} (\vb{k} + \vb{K}_-^{\pd}) \approx & \frac{t_1^2}{2 \, \sqrt{3} \, |t_2^{\pd}|} - \frac{3 \, \sqrt{3} \, |t_2^{\pd}|}{2} \left( k_y^3 - 3 \, k_y^{\pd} \, k_x^2 \right) + \mathcal{O} \left( k^4 \right).
\end{align}
Thus we have obtained a monkey saddle. Let us examine what happens when $M$ is close to but not equal to $-M_0^{\,}$, i.e $M = -M_0^{\,} + \delta M$. Assuming $t_2^{\,} > 0$, for very small $\delta M$, the Taylor expansion gets modified as 
\begin{align}
\varepsilon_2^{\pd} (\vb{k} + \vb{K}_-^{\pd}) \approx & \left| - \frac{t_1^2}{2 \, \sqrt{3} \, t_2^{\,}} + \delta M \right| + \frac{27 \, t_2^2 \,\, \delta M}{2 \, t_1^2} \left( k_x^2 + k_y^2 \right) \nonumber \\
& \quad - \left( \frac{3 \, \sqrt{3} \, t_2^{\,}}{2} + \frac{27 \, t_2^2 \,\, \delta M}{4 \, t_1^2} \right) \left( k_y^3 - 3 \, k_y^{\pd} \, k_x^2 \right) + \mathcal{O} \left( k^4, \delta M^2 \right).
\end{align}

As we see, a small quadratic term modulated by $\delta M$ appears in the dispersion, splitting the higher order critical point into an ordinary maximum/minimum at the centre, surrounded by three regular saddle points (see Fig~\ref{fig:Monkey}). Using the numerical values: $t = 1$ and $t = 1/2$:
\begin{align}
\varepsilon_2^{\pd} (\vb{k} + \vb{K}_-^{\pd}) \approx & \, \frac{1}{\sqrt{3}} - \delta M + \frac{27}{8} \, \delta M \left( k_x^2 + k_y^2 \right) \nonumber \\
& \quad - \left( \frac{3\sqrt{3}}{4} + \frac{27}{16} \, \delta M \right) \left( k_y^3 - 3 \, k_y^{\pd} \, k_x^2 \right) + \mathcal{O} \left( k^4, \delta M^2 \right).
\end{align}
We will now attempt to reproduce this Taylor expansion using the method described above. Firstly we calculate the eigenvalues and eigenvectors at $\vb{K}_-^{\,}$ for the \emph{critical} tuning (since $\delta M$ is treated here as a perturbation, we are computing eigenvalues and eigenvectors of $H(\vb{K}_-^{\,} + \lambda \, \vb{k} ; -M_0^{\,} + \lambda \, \delta M)$ at $\lambda = 0$, which is the same as computing them for the critical tuning). The Hamiltonian takes the value
\begin{equation}
\widehat{H}(\vb{K}_-^{\,}) = \begin{pmatrix}
-\frac{1}{\sqrt{3}} & 0 \\
0 & \frac{1}{\sqrt{3}} 
\end{pmatrix}
\end{equation}
Thus it is already diagonal. The eigenvalues and eigenvectors are then given by
\begin{align}
\varepsilon_{1}^{\,} (\vb{K}_-^{\,}) &= - \frac{1}{\sqrt{3}}, \quad |E_1^{\,} \rangle = \begin{pmatrix}
1 \\
0
\end{pmatrix}
,
\nonumber
\\
\varepsilon_{2}^{\,} (\vb{K}_-^{\,}) &= \frac{1}{\sqrt{3}}, \quad |E_2^{\,} \rangle = \begin{pmatrix}
0 \\
1
\end{pmatrix}
.
\end{align}
We now compute the matrices $\partial \widehat{H}$, $\partial^2 \widehat{H}$ and $\partial^3 \widehat{H}$, defined in Eq~\ref{eq:Hamltn_derivs}, in the eigenbasis:
\begin{subequations}
\begin{align}
\partial \widehat{H} =& \left(
\begin{array}{cc}
 \delta M & \frac{3}{2} i \left(k_x+i k_y\right) \\
 -\frac{3}{2} i \left(k_x-i k_y\right) & -\delta M \\
\end{array}
\right) 
\\
\partial^2 \widehat{H} =& \left(
\begin{array}{cc}
 \frac{9}{4} \sqrt{3} \left(k_x^2+k_y^2\right) & -\frac{3}{4} \left(k_x-i k_y\right){}^2
   \\
 -\frac{3}{4} \left(k_x+i k_y\right){}^2 & -\frac{9}{4} \sqrt{3} \left(k_x^2+k_y^2\right)
   \\
\end{array}
\right)
\\
\partial^3 \widehat{H} =& \left(
\begin{array}{cc}
 \frac{9}{8} \sqrt{3} k_y \left(k_y^2-3 k_x^2\right) & -\frac{9}{8} i \left(k_x-i
   k_y\right) \left(k_x+i k_y\right){}^2 \\
 \frac{9}{8} i \left(k_x-i k_y\right){}^2 \left(k_x+i k_y\right) & -\frac{9}{8} \sqrt{3}
   k_y \left(k_y^2-3 k_x^2\right) \\
\end{array}
\right)
\end{align}
\end{subequations}
The derivatives $\partial^N \varepsilon_2^{\,}$ for $N = 1, 2, 3$ are then given by
\begin{align}
\partial \varepsilon_2^{\,} =& \langle E_2^{\,} | \partial \widehat{H} | E_2^{\,} \rangle \nonumber 
\\
=& -\delta M, \nonumber 
\\
\partial^2 \varepsilon_2^{\,} =& \langle E_2^{\,} | \partial^2 \widehat{H} | E_2^{\,} \rangle + 2 \, \frac{\langle E_2^{\,} | \partial \widehat{H} | E_1^{\,} \rangle \langle E_1^{\,} | \partial \widehat{H} | E_2^{\,} \rangle}{\varepsilon_2^{\,} (\vb{K}_-^{\,}) - \varepsilon_1^{\,} (\vb{K}_-^{\,})} \nonumber \\
=& -\frac{9\sqrt{3}}{4} \left( k_x^2 + k_y^2 \right) + 2 \, \frac{-\frac{3 i}{2} \left(k_x^{\,} - i k_y^{\,} \right) \times \frac{3 i}{2} \left( k_x^{\,} + i k_y^{\,} \right)}{ \frac{1}{\sqrt{3}} - \left( - \frac{1}{\sqrt{3}} \right)} \nonumber \\
=& 0,
\nonumber \\
\partial^3 \varepsilon_n^{\,} =& \langle E_2^{\,} | \partial^3 \widehat{H} | E_2^{\,} \rangle + 3 \, \frac{\langle E_2^{\,} | \partial^2 \widehat{H} | E_1^{\,} \rangle \langle E_1^{\,} | \partial \widehat{H} | E_2^{\,} \rangle}{\varepsilon_2^{\,} (\vb{K}_-^{\,}) - \varepsilon_1^{\,} (\vb{K}_-^{\,})} + 3 \, \frac{\langle E_2^{\,} | \partial \widehat{H} | E_1^{\,} \rangle \langle E_1^{\,} | \partial^2 \widehat{H} | E_2^{\,} \rangle}{\varepsilon_2^{\,} (\vb{K}_-^{\,}) - \varepsilon_1^{\,} (\vb{K}_-^{\,})} \nonumber  
\\
& \quad + 6 \, \frac{\langle E_2^{\,} | \partial \widehat{H} | E_1^{\,} \rangle \langle E_1^{\,} | \partial \widehat{H} | E_1^{\,} \rangle \langle E_1^{\,} | \partial \widehat{H} | E_2^{\,} \rangle}{\left( \varepsilon_2^{\,} (\vb{K}_-^{\,}) - \varepsilon_1^{\,} (\vb{K}_-^{\,}) \right) \left( \varepsilon_2^{\,} (\vb{K}_-^{\,}) - \varepsilon_1^{\,} (\vb{K}_-^{\,}) \right)} \nonumber 
\\
& \quad - 6 \, \partial \varepsilon_2^{\,} \, \frac{\langle E_2^{\,} | \partial \widehat{H} | E_1^{\,} \rangle \langle E_1^{\,} | \partial \widehat{H} | E_2^{\,} \rangle}{\left( \varepsilon_2^{\,} (\vb{K}_-^{\,}) - \varepsilon_1^{\,} (\vb{K}_-^{\,}) \right)^2} \nonumber 
\\
=& - \frac{9 \sqrt{3}}{8} \left( k_y^3 - 3 \, k_x^2 k_y^{\,} \right) + 3 \, \frac{2 \, \text{Re} \left[ -\frac{3}{4} \left( k_x^{\,} + i k_y^{\,} \right)^2 \times \frac{3 i}{2} \left( k_x^{\,} + i \, k_y^{\,} \right) \right]}{\frac{1}{\sqrt{3}} - \left( - \frac{1}{\sqrt{3}} \right)} \nonumber 
\\
& \quad + 6 \, \frac{-\frac{3 i}{2} \left(k_x^{\,} - i k_y^{\,} \right) \times \delta M \times \frac{3 i}{2} \left( k_x^{\,} + i k_y^{\,} \right)}{ \left( \frac{1}{\sqrt{3}} - \left( - \frac{1}{\sqrt{3}} \right) \right)^2} \nonumber 
\\
& \quad - 6 (- \delta M) \, \frac{-\frac{3 i}{2} \left(k_x^{\,} - i k_y^{\,} \right) \times \frac{3 i}{2} \left( k_x^{\,} + i k_y^{\,} \right)}{ \left( \frac{1}{\sqrt{3}} - \left( - \frac{1}{\sqrt{3}} \right) \right)^2} \nonumber 
\\
= & - \frac{9 \sqrt{3}}{2} \left( k_y^3 - 3 \, k_x^2 k_y^{\,} \right) + \frac{81}{4} \, \delta M \left( k_x^2 + k_y^2 \right).
\end{align}
We can assemble these into the Taylor expansion using Eq~\ref{eq:Tayl_expansion_2}:
\begin{equation}
\varepsilon_2^{\pd} (\vb{k} + \vb{K}_-^{\pd}) \approx \, \frac{1}{\sqrt{3}} - \delta M + \frac{27}{8} \, \delta M \left( k_x^2 + k_y^2 \right) - \frac{3\sqrt{3}}{4} \left( k_y^3 - 3 \, k_y^{\pd} \, k_x^2 \right) + \cdots
\end{equation}

As it is evident, the Taylor expansion matches with the one we obtained by directly Taylor expanding the band dispersion, except for the extra $\delta M$ modulation of the coefficient of the cubic term. The reason why we did not obtain that term is that it arises from the expansion to fourth order in the auxiliary parameter $\lambda$ (basically from a term of the form $\lambda^4 \, \delta M \, \left( k_y^3 - 3 k_x^2 k_y \right)$), while here we have expanded only up to cubic order for the sake of brevity. Nevertheless, the main feature, which is the appearance of the small quadratic part that perturbs the monkey saddle when we tune off the critical value of the staggered chemical potential, is quite manifest here.

\section{Conclusion}\label{sec:conclusion}

In this paper, we set in detail the method to find HOVHS in practice, when the Hamiltonian $H(\vb{k})$ is known by various means. The knowledge of  $H(\vb{k})$ can come from tight-binding models or ab-initio calculations or even motivated by experimental results. The fundamental idea outlined above was to use an extension the Feynman Hellmann theorem to compute the various derivatives of the dispersion of any specified band using the technique explained in Sec~\ref{sec:method}. These derivatives were then used to assemble the Taylor expansion of the band to any specified order, at arbitrary $\vb{k}$ points, with the added possibility of incorporating the influence of tuning of any parameters in the Hamiltonian. The latter result is particularly timely, given the increasing interest and the remarkable recent progress in experimental methods to engineer exotic phases in materials by tuning them (via stress, twist, pressure, bias voltage, etc). Widely studied metallic systems like Sr$_2$RuO$_4$ or semiconducting ones like two dimensional InSe and In$_2$Se$_3$ as well as a wealth of other quantum materials, can be readily explored for the possibility to host HOVHS by this method. This is exactly where the fundamental computational method developed here, will demonstrate its full potential.

In the case of a non-trivial degeneracy at a point in k-space (such as band crossing, band touching etc.), the process of the computation of the derivatives and terms in the series expansion becomes more complicated. This is the subject of a future work and beyond the scope of the present. 

\paragraph{Note}
A readily usable implementation of the algorithm in Wolfram Mathematica is available~\cite{bandutils}.

\medskip
\textbf{Supporting Information} \par 
Supporting Information is available from the Wiley Online Library or from the author.

\medskip
\textbf{Acknowledgements} \par 
We would like to thank Claudio Chamon, Mark Greenaway, Nick Hine, Phil King, Samuel Magorrian, Edgar Abarca Morales, Andreas Rost and Peter Wahl for useful discussions. The work was supported by the EPSRC grant EP/T034351/1.

\begin{appendix}

\section{Derivation of the expressions}\label{sec:explicit_proof}
In this section, we explicitly prove the formula in Eq~\ref{eq:band_derivative_algorithm} for linear, quadratic and cubic derivatives of the band. For the sake of convenience of notation, we drop the `hats' on the Hamiltonian $\widehat{H}$ and write it simply as $H$. We further assume that all the eigenvectors are normalised, i.e $\langle n | n \rangle = 1$.

\subsection{First derivative}
The eigenvalue equation reads
\begin{equation}
H | n \rangle = \varepsilon_n^{\,} | n \rangle .
\end{equation}
Differentiating this once and applying the product rule for derivatives, we get 
\begin{equation}
\partial H | n \rangle + H | \partial n \rangle = \partial \varepsilon_n^{\,} | n \rangle + \varepsilon_n^{\,} | \partial n \rangle .
\label{eq:deriv_eig_val_eqn}
\end{equation}
(There is no problem with applying product rule since we are working with a finite dimensional vector space where all the terms in the eigenvalue equation are simply finite sums of product of vector and matrix coefficients in some basis). Now we take inner product of this equation with $| n \rangle $
\begin{align}
\langle n | \partial H | n \rangle + \langle n | H | \partial n \rangle &= \partial \varepsilon_n^{\,}  \langle n | n \rangle + \varepsilon_n^{\,} \langle n | \partial n \rangle , \nonumber \\
\implies \partial \varepsilon_n^{\,} &= \langle n | \partial H | n \rangle . \label{eq:band_deriv}
\end{align}
This is of course the familiar Feynman Hellmann result. We now take inner product of Eq~\ref{eq:deriv_eig_val_eqn} with another eigenstate $| m \rangle $ that is not degenerate to $| n \rangle$:
\begin{align}
\langle m | \partial H | n \rangle + \langle m | H | \partial n \rangle &= \partial \varepsilon_n^{\,}  \langle m | n \rangle + \varepsilon_n^{\,} \langle m | \partial n \rangle , \nonumber \\
\implies \langle m | \partial n \rangle &= \frac{\langle m | \partial H | n \rangle}{\varepsilon_n^{\,} - \varepsilon_m^{\,}} .
\label{eq:overlap_1st_deriv}
\end{align}
Thus, we can find the overlap of the derivative of the eigenvector $| n \rangle $, denoted by $| \partial n \rangle $ with the other eigenvectors \emph{not degenerate} to it. It is not possible within this framework to compute $\langle n | \partial n \rangle$ or even overlap with eigenvectors degenerate to it. We can however differentiate $\langle n | n \rangle = 1$ to obtain $\text{Re} \, \langle n | \partial n \rangle = 0$. The imaginary part is of course the famous Berry connection that can not be gauged out in general (specifically when we have topological bands). The fact that we can not precisely compute $\langle n | \partial n \rangle$ will not be an impediment to computing the higher derivatives of the band dispersion $\varepsilon_n^{\,} (\vb{k})$, as we shall show below. We also compute the overlap of Eq~\ref{eq:deriv_eig_val_eqn} with respect to another eigenstate that is degenerate to $|n \rangle$, denoted by $| n_1^{\,} \rangle \in D(n)$ (where $D(n)$ is the set of eigenvectors degenerate to $n$)
\begin{align}
\langle n_1^{\,} | \partial H | n \rangle + \langle n_1^{\,} | H | \partial n \rangle &= \partial \varepsilon_n^{\,}  \langle n_1^{\,} | n \rangle + \varepsilon_n^{\,} \langle n_1^{\,} | \partial n \rangle , \nonumber \\
\implies \langle n_1^{\,} | \partial H | n \rangle &= 0 , \,\, \text{for} \,\, \varepsilon_{n_1}^{\,} = \varepsilon_n^{\,} \,\, \text{but} \, |n_1^{\,} \rangle \neq | n \rangle .
\label{eq:overlap_1st_deriv_degen}
\end{align}
In practice however, we may find the projection of the first derivative of Hamiltonian matrix $\partial H$ to the degenerate subspace to be non-diagonal. That is, we may find $\langle n_1^{\,} | \partial H | n_2^{\,} \rangle \neq 0$, for distinct but degenerate $|n_1^{\,} \rangle$ and $|n_2^{\,} \rangle$. In such situations, we diagonalise the projection of $\partial H$ in this subspace, and replace the original set of $|n_i^{\,} \rangle$ with the new eigenbasis, so that Eq~\ref{eq:overlap_1st_deriv_degen} is automatically satisfied.

\subsection{Second derivative}
Differentiating Eq~\ref{eq:deriv_eig_val_eqn} once again we get
\begin{equation}
\partial^2 H | n \rangle + 2 \, \partial H | \partial n \rangle + H | \partial^2 n \rangle = \partial^2 \varepsilon_n^{\,} | n \rangle + 2 \, \partial \varepsilon_n^{\,} | \partial n \rangle + \varepsilon_n^{\,} | \partial^2 n \rangle .
\label{eq:2nd_deriv_eig_val_eqn}
\end{equation}
Taking inner product with $| n \rangle$, we get
\begin{equation}
 \langle n | \partial^2 H | n \rangle + 2 \, \langle n | \partial H | \partial n \rangle + \langle n | H | \partial^2 n \rangle = \partial^2 \varepsilon_n^{\,} \langle n | n \rangle + 2 \, \partial \varepsilon_n^{\,} \, \langle n | \partial n \rangle + \varepsilon_n^{\,} \langle n | \partial^2 n \rangle
\end{equation}
On rearranging terms, inserting a resolution of identity and splitting into projectors over the degenerate and non-degenerate subspaces, this becomes 
\begin{align}
\partial^2 \varepsilon_n^{\,} =& \langle n | \partial^2 H | n \rangle + 2 \, \langle n | \partial H | \partial n \rangle - 2 \, \partial \varepsilon_n^{\,} \, \langle n | \partial n \rangle \nonumber \\
 =& \langle n | \partial^2 H | n \rangle + 2 \sum_{| m \rangle \notin D(n)} \langle n | \partial H | m \rangle \langle m | \partial n \rangle + 2 \sum_{|n_1^{\,} \rangle \in D(n)} \langle n | \partial H | n_1^{\,} \rangle \langle n_1^{\,} | \partial n \rangle - 2 \, \partial \varepsilon_n^{\,} \, \langle n | \partial n \rangle \nonumber \\
  =& \langle n | \partial^2 H | n \rangle + 2 \sum_{| m \rangle \notin D(n)} \frac{\langle n | \partial H | m \rangle \langle m | \partial H | n \rangle}{\varepsilon_n^{\,} - \varepsilon_m^{\,}},
\label{eq:band_2nd_deriv}
\end{align}
where we have used $\langle n | \partial H | n \rangle = \partial \varepsilon_n^{\,}$ and $\langle n | \partial H | n_1^{\,} \rangle = 0$ when $| n_1^{\,} \rangle \neq | n \rangle$ in going to the last line. We now compute the overlap of the terms in Eq~\ref{eq:2nd_deriv_eig_val_eqn} with $| m \rangle \notin D(n)$:
\begin{equation}
\langle m_1^{\,} | \partial^2 H | n \rangle + 2 \, \langle m_1^{\,} | \partial H | \partial n \rangle + \langle m_1^{\,} | H | \partial^2 n \rangle = \partial^2 \varepsilon_n^{\,} \langle m_1^{\,} | n \rangle + 2 \, \partial \varepsilon_n^{\,} \langle m_1^{\,} | \partial n \rangle + \varepsilon_n^{\,} \langle m_1^{\,} | \partial^2 n \rangle .
\end{equation}
Simplifying this further by rearranging terms and introducing another resolution of identity:
\begin{align}
\langle m_1^{\,} | \partial^2 n \rangle = & \frac{1}{\varepsilon_n^{\,} - \varepsilon_{m_1}^{\,}} \big[ \langle m_1^{\,} | \partial^2 H | n \rangle + 2 \, \langle m_1^{\,} | \partial H | \partial n \rangle - 2 \, \partial \varepsilon_n^{\,} \langle m_1^{\,} | \partial n \rangle \big] \nonumber \\
=& \frac{1}{\varepsilon_n^{\,} - \varepsilon_{m_1}^{\,}} \bigg[ \langle m_1^{\,} | \partial^2 H | n \rangle + 2 \sum_{|m_2\rangle \notin D(n)} \langle m_1^{\,} | \partial H | m_2^{\,} \rangle \langle m_2^{\,} | \partial n \rangle \nonumber 
\\
& \quad + 2 \sum_{| n_1^{\,} \rangle \in D(n)} \langle m_1^{\,} | \partial H | n_1^{\,} \rangle \langle n_1^{\,} | \partial n \rangle - 2 \, \partial \varepsilon_n^{\,} \langle m_1^{\,} | \partial n \rangle \bigg] \nonumber
\\
= & \frac{\langle m_1^{\,} | \partial^2 H | n \rangle}{\varepsilon_n^{\,} - \varepsilon_{m_1}^{\,}} + 2 \sum_{|m_2\rangle \notin D(n)} \frac{\langle m_1^{\,} | \partial H | m_2^{\,} \rangle \langle m_2^{\,} | \partial H | n \rangle}{(\varepsilon_n^{\,} - \varepsilon_{m_1}^{\,})(\varepsilon_n^{\,} - \varepsilon_{m_2}^{\,})} - 2 \, \partial \varepsilon_n^{\,} \frac{\langle m_1^{\,} | \partial H | n \rangle}{(\varepsilon_n^{\,} - \varepsilon_{m_1}^{\,})^2} \nonumber 
\\
& \quad + 2 \sum_{n_1^{\,} \in D(n)} \frac{\langle m_1^{\,} | \partial H | n_1^{\,} \rangle}{\varepsilon_n^{\,} - \varepsilon_{m_1}^{\,}} \langle n_1^{\,} | \partial n \rangle .
\label{eq:overlap_2nd_deriv}
\end{align}
This will be used in the computation of the third derivative below. Note that once again we are not able to explicitly evaluate the overlap with $| n \rangle$. Lastly, we compute the overlap of Eq~\ref{eq:band_2nd_deriv} with $| n_1^{\,} \rangle \in D(n)$ and $| n_1^{\,} \rangle \neq |n \rangle$
\begin{equation}
\langle n_1^{\,} | \partial^2 H | n \rangle + 2 \, \langle n_1^{\,} | \partial H | \partial n \rangle + \langle n_1^{\,} | H | \partial^2 n \rangle = \partial^2 \varepsilon_n^{\,} \langle n_1^{\,} | n \rangle + 2 \, \partial \varepsilon_n^{\,} \langle n_1^{\,} | \partial n \rangle + \varepsilon_n^{\,} \langle n_1^{\,} | \partial^2 n \rangle .
\end{equation}
Using $\langle n_1^{\,} | H = \varepsilon_n^{\,} \langle n_1^{\,} |$, introducing resolution of identity and simplifying
\begin{align}
\langle n_1^{\,} | \partial^2 H | n \rangle + 2 \sum_{|m_1 \rangle \notin D(n)} \langle n_1^{\,} | \partial H | m_1^{\,} \rangle \langle m_1^{\,} | \partial n \rangle + 2 \sum_{|n_2 \rangle \notin D(n)} \langle n_1^{\,} | \partial H | n_2^{\,} \rangle \langle n_2^{\,} | \partial n \rangle = & 2 \, \partial \varepsilon_n^{\,} \langle n_1^{\,} | \partial n \rangle , \nonumber \\
 \implies \langle n_1^{\,} | \partial^2 H | n \rangle + 2 \sum_{|m_1 \rangle \notin D(n)} \frac{\langle n_1^{\,} | \partial H | m_1^{\,} \rangle \langle m_1^{\,} | \partial H | n \rangle}{\varepsilon_n^{\,} - \varepsilon_{m_1}^{\,}}  = & 0. \label{eq:second_deriv_overlap_nn}
\end{align}
The expression on the left hand side resembles the expression for $\partial^2 \varepsilon_n^{\,}$, with the difference that the leftmost bra in each of the terms is $|n_1^{\,}\rangle$ instead of $|n\rangle$, and the expression ultimately evaluates to $0$ rather than $\partial^2 \varepsilon_n^{\,}$.

\subsection{Third derivative}
The third derivative of the eigenvalue equation reads
\begin{equation}
\partial^3 H | n \rangle + 3 \, \partial^2 H | \partial n \rangle + 3 \, \partial H | \partial^2 n \rangle + H | \partial^3 n \rangle 
= \partial^3 \varepsilon_n^{\,} | n \rangle + 3 \, \partial^2 \varepsilon_n^{\,} | \partial n \rangle + 3 \, \partial \varepsilon_n^{\,} | \partial^2 n \rangle + \varepsilon_n^{\,} | \partial^3 n \rangle .
\label{eq:3rd_deriv_eig_val_eqn}
\end{equation}
We compute the overlap of this equation with $| n \rangle$, rearrange terms, introduce resolutions of identity (splitting them into projectors over the degenerate and non-degenerate subspaces) and do some simplifications
\begin{align}
\partial^3 \varepsilon_n^{\,} =& \langle n | \partial^3 H | n \rangle + 3 \langle n | \partial^2 H | \partial n \rangle + 3 \langle n | \partial H | \partial^2 n \rangle - 3 \, \partial^2 \varepsilon_n^{\,} \, \langle n | \partial n \rangle - 3 \, \partial \varepsilon_n^{\,} \, \langle n | \partial^2 n \rangle \nonumber 
\\ 
=& \langle n | \partial^3 H | n \rangle + 3 \sum_{|m_1 \rangle \notin D(n)} \langle n | \partial^2 H | m_1^{\,} \rangle \langle m_1^{\,} | \partial n \rangle + 3 \sum_{|m_1 \rangle \notin D(n)} \langle n | \partial H | m_1^{\,} \rangle \langle m_1^{\,} | \partial^2 n \rangle \nonumber 
\\
& \quad + 3 \sum_{| n_1^{\,} \rangle \in D(n)} \langle n | \partial^2 H | n_1^{\,} \rangle \langle n_1^{\,} | \partial n \rangle + 3 \sum_{| n_1^{\,} \rangle \in D(n)} \langle n | \partial H | n_1^{\,} \rangle \langle n_1^{\,} | \partial^2 n \rangle \nonumber \\
& \quad - 3 \,\, \partial^2 \varepsilon_n^{\,} \, \langle n | \partial n \rangle - 3 \,\, \partial \varepsilon_n^{\,} \, \langle n | \partial^2 n \rangle .
\end{align}
We can substitute for $\langle m_1^{\,} | \partial n \rangle$ and $\langle m_1^{\,} | \partial^2 n \rangle$ from Eqs~\ref{eq:overlap_1st_deriv} and \ref{eq:overlap_2nd_deriv} respectively to obtain

\begin{align}
\partial^3 \varepsilon_n^{\,} 
=& \langle n | \partial^3 H | n \rangle + 3 \sum_{|m_1 \rangle \notin D(n)} \frac{\langle n | \partial^2 H | m_1^{\,} \rangle \langle m_1^{\,} | \partial H | n \rangle}{\varepsilon_n^{\,} - \varepsilon_{m_1}^{\,}} \nonumber 
\\
& \quad + 3 \sum_{|m_1 \rangle \notin D(n)} \frac{\langle n | \partial H | m_1^{\,} \rangle \langle m_1^{\,} | \partial^2 H | n \rangle}{\varepsilon_n^{\,} - \varepsilon_{m_1}^{\,}} \nonumber
\\
& \quad + 6 \sum_{|m_1 \rangle \notin D(n)} \sum_{|m_2 \rangle \notin D(n)} \frac{\langle n | \partial H | m_1^{\,} \rangle \langle m_1^{\,} | \partial H | m_2^{\,} \rangle \langle m_2^{\,} | \partial H | n \rangle}{(\varepsilon_n^{\,} - \varepsilon_{m_1}^{\,})(\varepsilon_n^{\,} - \varepsilon_{m_2}^{\,})} \nonumber 
\\
& \quad -6 \,\, \partial \varepsilon_n^{\,} \sum_{|m_1 \rangle \notin D(n)} \frac{\langle n | \partial H | m_1^{\,} \rangle \langle m_1^{\,} | \partial H | n \rangle}{(\varepsilon_n^{\,} - \varepsilon_{m_1}^{\,})^2} \nonumber
\\
& \quad + (*1) \,\, 6 \sum_{| n_1^{\,} \rangle \in D(n)} \sum_{| m_1 \rangle \notin D(n)} \frac{\langle n | \partial H | m_1^{\,} \rangle \langle m_1^{\,} | \partial H | n_1^{\,} \rangle}{\varepsilon_n^{\,} - \varepsilon_{m_1}^{\,}} \langle n_1^{\,} | \partial n \rangle \nonumber 
\\
& \quad + (*2) \,\, 3 \sum_{| n_1^{\,} \rangle \in D(n)} \langle n | \partial^2 H | n_1^{\,} \rangle \langle n_1^{\,} | \partial n \rangle \nonumber \\
& \quad + (*3) \,\, 3 \sum_{| n_1^{\,} \rangle \in D(n)} \langle n | \partial H | n_1^{\,} \rangle \langle n_1^{\,} | \partial^2 n \rangle \nonumber \\
& \quad  - 3 \,\, \partial^2 \varepsilon_n^{\,} \, \langle n | \partial n \rangle - 3 \,\, \partial \varepsilon_n^{\,} \, \langle n | \partial^2 n \rangle . \label{eq:band_third_deriv_intermediate}
\end{align}
Now let us look at some of the terms marked with a *. To simplify them, we pull out the term containing $|n\rangle$ from the sum over $|n_1^{\,} \rangle \in D(n)$ and substitute the expression for $\partial^2 \varepsilon_n^{\,}$:
\begin{align}
(*2) + (&*1) \nonumber \\
=& \, 3 \sum_{| n_1^{\,} \rangle \in D(n)} \langle n | \partial^2 H | n_1^{\,} \rangle \langle n_1^{\,} | \partial n \rangle + 6 \sum_{| n_1^{\,} \rangle \in D(n)} \sum_{| m_1 \rangle \notin D(n)} \frac{\langle n | \partial H | m_1^{\,} \rangle \langle m_1^{\,} | \partial H | n_1^{\,} \rangle}{\varepsilon_n^{\,} - \varepsilon_{m_1}^{\,}} \langle n_1^{\,} | \partial n_1^{\,} \rangle \nonumber \\
=& \, 3 \left( \langle n | \partial^2 H | n \rangle + 2 \sum_{| m_1 \rangle \notin D(n)} \frac{\langle n | \partial H | m_1^{\,} \rangle \langle m_1^{\,} | \partial H | n \rangle}{\varepsilon_n^{\,} - \varepsilon_{m_1}^{\,}} \right) \langle n | \partial n \rangle \nonumber \\
& + 3 \sum_{\substack{|n_1^{\,} \rangle \in D(n) \\ | n_1^{\,} \rangle \neq | n \rangle}} \bigg[ \langle n | \partial^2 H | n_1^{\,} \rangle + 2 \sum_{| m_1 \rangle \notin D(n)} \frac{\langle n | \partial H | m_1^{\,} \rangle \langle m_1^{\,} | \partial H | n_1^{\,} \rangle}{\varepsilon_n^{\,} - \varepsilon_{m_1}^{\,}} \bigg] \langle n_1^{\,} | \partial n_1^{\,} \rangle \nonumber \\
=&  \, 3 \,\, \partial^2 \varepsilon_n^{\,} \, \langle n | \partial n \rangle,
\end{align}
where we substitute from Eq~\ref{eq:band_2nd_deriv} for the terms within parenthesis, while the terms within [ ] vanish due to Eq~\ref{eq:second_deriv_overlap_nn} (since $|n_1^{\,} \rangle \neq |n\rangle$). Lastly, since $\langle n | \partial H | n \rangle = \partial \varepsilon_n^{\,}$ (from Eq~\ref{eq:band_deriv}) and $\langle n | \partial H | n_1^{\,} \rangle = 0$ for $| n \rangle \neq | n_1^{\,} \rangle$ (from Eq~\ref{eq:overlap_1st_deriv_degen}), we have
\begin{equation}
(*3) = 3 \sum_{| n_1^{\,} \rangle \in D(n)} \langle n | \partial H | n_1^{\,} \rangle \langle n_1^{\,} | \partial^2 n \rangle = 3 \,\, \partial \varepsilon_n^{\,} \, \langle n | \partial^2 n \rangle.
\end{equation}
Therefore the sum $(*1) + (*2) + (*3)$ cancels exactly with the last line of Eq~\ref{eq:band_third_deriv_intermediate} so that the expression for the third derivative of the band becomes
\begin{align}
\partial^3 \varepsilon_n^{\,} 
=& \langle n | \partial^3 H | n \rangle + 3 \sum_{|m_1 \rangle \notin D(n)} \frac{\langle n | \partial^2 H | m_1^{\,} \rangle \langle m_1^{\,} | \partial H | n \rangle}{\varepsilon_n^{\,} - \varepsilon_{m_1}^{\,}} + 3 \sum_{|m_1 \rangle \notin D(n)} \frac{\langle n | \partial H | m_1^{\,} \rangle \langle m_1^{\,} | \partial^2 H | n \rangle}{\varepsilon_n^{\,} - \varepsilon_{m_1}^{\,}} \nonumber
\\
& \quad + 6 \sum_{|m_1 \rangle \notin D(n)} \sum_{|m_2 \rangle \notin D(n)} \frac{\langle n | \partial H | m_1^{\,} \rangle \langle m_1^{\,} | \partial H | m_2^{\,} \rangle \langle m_2^{\,} | \partial H | n \rangle}{(\varepsilon_n^{\,} - \varepsilon_{m_1}^{\,})(\varepsilon_n^{\,} - \varepsilon_{m_2}^{\,})} \nonumber 
\\
& \quad -6 \,\, \partial \varepsilon_n^{\,} \sum_{|m_1 \rangle \notin D(n)} \frac{\langle n | \partial H | m_1^{\,} \rangle \langle m_1^{\,} | \partial H | n \rangle}{(\varepsilon_n^{\,} - \varepsilon_{m_1}^{\,})^2}.
\label{eq:band_3rd_deriv}
\end{align}
To obtain this, we got rid of the terms containing $\langle n | \partial n \rangle$ and $\langle n | \partial^2 n \rangle$ by a careful application of the results obtained for the lower derivatives $\partial \varepsilon_n^{\,}$ and $\partial^2 \varepsilon_n^{\,}$, along with the expressions involving the overlaps $\langle n_1^{\,} | \partial H | n \rangle$ and $\langle n_1^{\,} | \partial^2 H | n \rangle$ for $|n_1^{\,} \rangle \in D(n)$ and $|n_1^{\,} \rangle \neq | n \rangle$.

\section{General proof}
In this section we prove the general case by induction. Since the proof is somewhat involved, we provide a brief outline of it. While computing the expression for $\partial^N \varepsilon_n^{\,}$, we will encounter terms like $\langle m_1^{\,} | \partial^M m_2^{\,} \rangle$, which can not be evaluated in general. The strategy will be to expand such terms (particularly when $|m_1^{\,} \rangle$ and $|m_2^{\,} \rangle$ are non-degenerate), substitute the expansions in the expression for $\partial^N \varepsilon_n^{\,}$, and get the pathological terms to cancel out. We first apply the Leibniz theorem for higher derivatives of products to differentiate the eigenvalue equation:
\begin{subequations}
\begin{align}
\partial^{N} (H \, |n \rangle) &= \partial^{N} (\varepsilon_n^{\,} \, |n \rangle) \\
\implies \sum_{K=0}^{N} \begin{pmatrix}
N \\
K
\end{pmatrix} \partial^{N-K} H \, | \partial^{K} n \rangle &= \sum_{K=0}^{N} \begin{pmatrix}
N \\
K
\end{pmatrix} \partial^{N-K} \varepsilon_n^{\,} \, | \partial^{K} n \rangle 
\end{align}
\end{subequations}
Isolating the $K = 0$ and $K = N$ terms on both sides of the equation, we obtain
\begin{equation}
H \, |\partial^{N} n \rangle + \partial^N H \, |n \rangle + \sum_{K=1}^{N-1} \begin{pmatrix}
N \\
K
\end{pmatrix} \partial^{N-K} H \, | \partial^{K} n \rangle 
= \varepsilon_n^{\,} \, |\partial^{N} n \rangle + \partial^N \varepsilon_n^{\,} \, |n \rangle + \sum_{K=1}^{N-1} \begin{pmatrix}
N \\
K
\end{pmatrix} \partial^{N-K} \varepsilon_n^{\,} \, | \partial^{K} n \rangle \label{eq:nth_deriv_eig_eqn}
\end{equation}
After taking inner product on both sides with $|m\rangle \notin D(n)$ (i.e an eigenstate not degenerate to $|n\rangle$), using $\langle m | H = \varepsilon_m^{\,} \, \langle m|$ and rearranging, we obtain:
\begin{equation}
\langle m | \partial^{N} n \rangle =  
 \frac{1}{\varepsilon_n^{\,} - \varepsilon_m^{\,}} \left[ \langle m | \partial^N H | n \rangle + \sum_{K = 1}^{N-1} \begin{pmatrix}
N \\
K
\end{pmatrix}
\left(
\langle m | \partial^{N-K} H ** | \partial^{K} n \rangle - \partial^{N-K} \varepsilon_n^{\,} \, \langle m | \partial^{K} n \rangle \right) \right]
\end{equation}

We now introduce a resolution of identity in terms of the eigenvectors of the Hamiltonian at the given $\vb{k}_0^{\,}$ at the location indicated by ** and separate out the projector  to the degenerate subspace $\sum_{|\tilde{n}\rangle \in D(n)} | \tilde{n} \rangle \langle \tilde{n} |$. For convenience, we add subscripts to the variables and change $N \rightarrow K_i$ so that the expression becomes more readily usable in the forthcoming calculations:
\begin{multline}
\langle m_i^{\,} | \partial^{K_i} n \rangle = \frac{1}{\varepsilon_n^{\,} - \varepsilon_{m_i}^{\,}} \bigg[ \langle m_i^{\,} | \partial^{K_i} H | n \rangle + \sum_{K_{i+1} = 1}^{K_i-1} \begin{pmatrix}
K_i \\
K_{i+1}
\end{pmatrix}
\sum_{|m_{i+1} \rangle \notin D(n)}
\bigg\{ 
\\
\left(
\langle m_i^{\,} | \partial^{K_i - K_{i+1}} H | m_{i+1}^{\,} \rangle - \partial^{K_i - K_{i+1}} \varepsilon_n^{\,} \, \delta_{m_i, m_{i+1}} \right) \langle m_{i+1} | \partial^{K_{i+1}} n \rangle
\bigg\}
\\ 
+ \sum_{K_{i+1} = 1}^{K_i - 1} \begin{pmatrix}
K_i \\
K_{i+1}
\end{pmatrix}
\sum_{|\tilde{n} \rangle \in D(n)}
\langle m_i^{\,} | \partial^{K_i - K_{i+1}} H | \tilde{n} \rangle \langle \tilde{n} | \partial^{K_{i+1}} n \rangle \bigg]
\label{eq:overlap_less_compact}
\end{multline}

To make the notation less cumbersome, for a given $|n \rangle$, we define
\begin{subequations}
\begin{align}
& \widetilde{\mathcal{H}}_{m_i, m_{j}}^{K} \coloneqq  \langle m_i^{\,} | \partial^{K} H | m_{j}^{\,} \rangle - \partial^{K} \varepsilon_n^{\,} \, \delta_{m_i, m_{j}} \\
& \mathcal{H}_{m_i,m_j}^{K} \coloneqq \langle m_i^{\,} | \partial^{K} H | m_j^{\,} \rangle \\
& \mathcal{O}_{m_i, m_j}^{K} \coloneqq \langle m_i^{\,} | \partial^{K} m_j^{\,} \rangle \text{, for non-degenerate } | m_i^{\,} \rangle , | m_j^{\,} \rangle \\
& \mathcal{R}_{\tilde{n},n}^{K} \coloneqq \langle \tilde{n} | \partial^{K} n \rangle \text{, for degenerate } | \tilde{n} \rangle , |n\rangle.
\end{align}
\end{subequations}
where $\mathcal{O}$ is the overlap which will be represented as a recursive function, while the matrix elements $\widetilde{\mathcal{H}}$ and $\mathcal{H}$ will occur respectively in the middle and at the ends in each of the terms in the final expression for the band derivatives. The reason for separately defining the overlap $\mathcal{R}$, which occurs at the end of the sequence of product of terms, is that we can not define a recursive relationship for it unlike $\mathcal{O}$, and all of the terms containing $\mathcal{R}'$s will be shown to cancel out in the final expressions. We can now write the expression for the overlap in Eq~\ref{eq:overlap_less_compact} in a compact form:
\begin{multline}
\mathcal{O}_{m_i,n}^{K_i} = \frac{1}{\varepsilon_n^{\,} - \varepsilon_{m_i}^{\,}} \bigg[ \mathcal{H}_{m_i,n}^{K_i} + \sum_{K_{i+1} = 1}^{K_i-1} \begin{pmatrix}
K_i \\
K_{i+1}
\end{pmatrix}
\sum_{|m_{i+1} \rangle \notin D(n)} \widetilde{\mathcal{H}}_{m_i, m_{i+1}}^{K_i - K_{i+1}} \mathcal{O}_{m_{i+1},n}^{K_{i+1}} \\
 + \sum_{K_{i+1} = 1}^{K_i-1} \begin{pmatrix}
K_i \\
K_{i+1}
\end{pmatrix}
\sum_{|\tilde{n} \rangle \in D(n)}
\mathcal{H}_{m_i,\tilde{n}}^{K_i - K_{i+1}} \, \mathcal{R}_{\tilde{n},n}^{K_{i+1}} \bigg]
\label{eq:overlap_recursive}
\end{multline}
This clearly defines a recursive relationship between the overlaps $\mathcal{O}$. We now expand this by applying the analogous expression for $\mathcal{O}_{m_{i+1}, K_{i+1}}$:
\begin{align}
\mathcal{O}_{m_i,n}^{K_i} = & \frac{\mathcal{H}_{m_i,n}^{K_i}}{\varepsilon_n^{\,} - \varepsilon_{m_i}^{\,}} + 
\sum_{K_{i+1} = 1}^{K_i-1} \begin{pmatrix}
K_i \\
K_{i+1}
\end{pmatrix}
\sum_{|m_{i+1} \rangle \notin D(n)} \frac{\widetilde{\mathcal{H}}_{m_i, m_{i+1}}^{K_i - K_{i+1}}}{\varepsilon_n^{\,} - \varepsilon_{m_i}^{\,}} \bigg[ \frac{\mathcal{H}_{m_{i+1},n}^{K_{i+1}}}{\varepsilon_n^{\,} - \varepsilon_{m_{i+1}}^{\,}} \nonumber \\ 
& \qquad + \sum_{K_{i+2} = 1}^{K_{i+1}-1} \begin{pmatrix}
K_{i+1} \\
K_{i+2}
\end{pmatrix}
\sum_{|m_{i+2} \rangle \notin D(n)} \frac{\widetilde{\mathcal{H}}_{m_{i+1}, m_{i+2}}^{K_{i+1} - K_{i+2}}}{\varepsilon_n^{\,} - \varepsilon_{m_{i+1}}^{\,}} \, \mathcal{O}_{m_{i+2},n}^{K_{i+2}} \nonumber \\
& \qquad + \sum_{K_{i+2} = 1}^{K_{i+1}-1} \begin{pmatrix}
K_{i+1} \\
K_{i+2}
\end{pmatrix}
\sum_{|\tilde{n} \rangle \in D(n)}
\frac{\mathcal{H}_{m_{i+1},\tilde{n}}^{K_{i+1} - K_{i+2}}}{\varepsilon_n^{\,} - \varepsilon_{m_{i+1}}^{\,}} \, \mathcal{R}_{\tilde{n},n}^{K_{i+2}} \bigg] \nonumber \\
& \qquad + (*) \sum_{K_{i+1} = 1}^{K_i-1} \begin{pmatrix}
K_i \\
K_{i+1}
\end{pmatrix}
\sum_{|\tilde{n} \rangle \in D(n)}
\frac{\mathcal{H}_{m_i, \tilde{n}}^{K_i - K_{i+1}}}{\varepsilon_n^{\,} - \varepsilon_{m_i}^{\,}} \, \mathcal{R}_{\tilde{n},n}^{K_{i+1}} .
\end{align}

Noticeably, the $\widetilde{\mathcal{H}}$'s always occur at the beginning and in the middle of the products while both $\mathcal{H}$ and $\mathcal{R}$ terminate the sequence of products (since the $\mathcal{R}$'s can not be expanded further). We rearrange the above equation to collect the terms with $\mathcal{R}$'s together (after the $(*)$).

\begin{align}
\mathcal{O}_{m_i,n}^{K_i} = & \frac{\mathcal{H}_{m_i,n}^{K_i}}{\varepsilon_n^{\,} - \varepsilon_{m_i}^{\,}} + \sum_{K_{i+1} = 1}^{K_i-1} \begin{pmatrix}
K_i \\
K_{i+1}
\end{pmatrix}
\sum_{|m_{i+1} \rangle \notin D(n)} \frac{\widetilde{\mathcal{H}}_{m_i, m_{i+1}}^{K_i - K_{i+1}}}{\varepsilon_n^{\,} - \varepsilon_{m_i}^{\,}} 
\,
\frac{\mathcal{H}_{m_{i+1},n}^{K_{i+1}}}{\varepsilon_n^{\,} - \varepsilon_{m_{i+1}}^{\,}} \nonumber \\
& + \sum_{K_{i+1} = 1}^{K_i-1} \begin{pmatrix}
K_i \\
K_{i+1}
\end{pmatrix}
\sum_{K_{i+2} = 1}^{K_{i+1}-1} \begin{pmatrix}
K_{i+1} \\
K_{i+2}
\end{pmatrix}
\sum_{|m_{i+1} \rangle \notin D(n)}
\sum_{|m_{i+2} \rangle \notin D(n)}
\frac{\widetilde{\mathcal{H}}_{m_i, m_{i+1}}^{K_i - K_{i+1}}}{\varepsilon_n^{\,} - \varepsilon_{m_i}^{\,}}
\,
\frac{\widetilde{\mathcal{H}}_{m_{i+1}, m_{i+2}}^{K_{i+1} - K_{i+2}}}{\varepsilon_n^{\,} - \varepsilon_{m_{i+1}}^{\,}} \, \mathcal{O}_{m_{i+2},n}^{K_{i+2}} 
\nonumber \\
& +(*) \sum_{K_{i+1} = 1}^{K_i-1} \begin{pmatrix}
K_i \\
K_{i+1}
\end{pmatrix}
\sum_{|\tilde{n} \rangle \in D(n)}
\frac{\mathcal{H}_{m_i,\tilde{n}}^{K_i - K_{i+1}}}{\varepsilon_n^{\,} - \varepsilon_{m_i}^{\,}} \, \mathcal{R}_{\tilde{n},n}^{K_{i+1}}
\nonumber \\
& + \sum_{K_{i+1} = 1}^{K_i-1} \begin{pmatrix}
K_i \\
K_{i+1}
\end{pmatrix}
\sum_{K_{i+2} = 1}^{K_{i+1}-1} \begin{pmatrix}
K_{i+1} \\
K_{i+2}
\end{pmatrix}
\sum_{|m_{i+1} \rangle \notin D(n)}
\sum_{|\tilde{n} \rangle \in D(n)}
\frac{\widetilde{\mathcal{H}}_{m_i, m_{i+1}}^{K_i - K_{i+1}}}{\varepsilon_n^{\,} - \varepsilon_{m_i}^{\,}}
\,
\frac{\mathcal{H}_{m_{i+1},\tilde{n}}^{K_{i+1} - K_{i+2}}}{\varepsilon_n^{\,} - \varepsilon_{m_{i+1}}^{\,}} \, \mathcal{R}_{\tilde{n},n}^{K_{i+2}} .
\end{align}

We can expand this further by substituting the expression for $\mathcal{O}_{m_{i+2},n}^{K_{i+2}}$ based on Eq~\ref{eq:overlap_recursive}. That can be further expanded by substituting the expression for $\mathcal{O}_{m_{i+3},n}^{K_{i+3}}$, once again based on Eq~\ref{eq:overlap_recursive}. We cannot however continue indefinitely expanding like this. The reason is that we have $1 \leqslant K_{j+1} \leqslant K_{j} - 1$, which, on repeated application, starting from $K_{i+l}$ up to $K_{i}$, gives $1 \leqslant K_{i+l} \leqslant K_{i+l-1} - 1 \leqslant K_{i+l-2} - 2 \leqslant \cdots \leqslant K_{i} - l$. This means that the maximum allowed $l$ itself equals $K_{i} - 1$ (since $1 \leqslant K_{i+l} \leqslant K_i - l$). This also gives the upper limit of the recursive expansion. The full expanded form then reads

\begin{align}
\mathcal{O}_{m_i,n}^{K_i} = & \frac{\mathcal{H}_{m_i,n}^{K_i}}{\varepsilon_n^{\,} - \varepsilon_{m_i}^{\,}} + \sum_{K_{i+1} = 1}^{K_i-1} 
\sum_{|m_{i+1} \rangle \notin D(n)} 
\begin{pmatrix}
K_i \\
K_{i+1}
\end{pmatrix}
\frac{\widetilde{\mathcal{H}}_{m_i, m_{i+1}}^{K_i - K_{i+1}}}{\varepsilon_n^{\,} - \varepsilon_{m_i}^{\,}}
\, 
\frac{\mathcal{H}_{m_{i+1},n}^{K_{i+1}}}{\varepsilon_n^{\,} - \varepsilon_{m_{i+1}}^{\,}} \nonumber \\
& + \sum_{K_{i+1} = 1}^{K_i-1} \sum_{K_{i+2} = 1}^{K_{i+1}-1} 
\sum_{|m_{i+1} \rangle \notin D(n)} \sum_{|m_{i+2} \rangle \notin D(n)}
\begin{pmatrix}
K_i \\
K_{i+1}
\end{pmatrix}
\begin{pmatrix}
K_{i+1} \\
K_{i+2}
\end{pmatrix}
\frac{\widetilde{\mathcal{H}}_{m_i, m_{i+1}}^{K_i - K_{i+1}}}{\varepsilon_n^{\,} - \varepsilon_{m_i}^{\,}}
\,
\frac{\widetilde{\mathcal{H}}_{m_{i+1}, m_{i+2}}^{K_{i+1} - K_{i+2}}}{\varepsilon_n^{\,} - \varepsilon_{m_{i+1}}^{\,}}
\,
\frac{\mathcal{H}_{m_{i+2},n}^{K_{i+2}}}{\varepsilon_n^{\,} - \varepsilon_{m_{i+2}}^{\,}} \nonumber \\
& + \cdots \nonumber \\
& + \cdots \nonumber \\
& + \sum_{K_{i+1} = 1}^{K_i - 1} \sum_{K_{i+2} = 1}^{K_{i+1} - 1} \cdots \sum_{K_{i+l} = 1}^{K_{i+l-1} - 1} \sum_{|m_{i+1} \rangle \neq |n \rangle} \sum_{|m_{i+2} \rangle \neq |n \rangle} \cdots \sum_{|m_{i+l} \rangle \neq |n \rangle} \bigg[ \begin{pmatrix}
K_i \\
K_{i+1}
\end{pmatrix} 
\begin{pmatrix}
K_{i+1} \\
K_{i+2}
\end{pmatrix}
\cdots 
\begin{pmatrix}
K_{i+l-1} \\
K_{i+l}
\end{pmatrix} \nonumber \\
& \times \frac{\widetilde{\mathcal{H}}_{m_i, m_{i+1}}^{K_i - K_{i+1}}}{\varepsilon_n^{\,} - \varepsilon_{m_i}^{\,}} \times \frac{\widetilde{\mathcal{H}}_{m_{i+1}, m_{i+2}}^{K_{i+1} - K_{i+2}}}{\varepsilon_n^{\,} - \varepsilon_{m_{i+1}}^{\,}} \times \cdots \times \frac{\widetilde{\mathcal{H}}_{m_{i+l-1}, m_{i+l}}^{K_{i+l-1} - K_{i+l}}}{\varepsilon_n^{\,} - \varepsilon_{m_{i+l-1}}^{\,}}
\,
\frac{\mathcal{H}_{m_{i+l},n}^{K_{i+l}}}{\varepsilon_n^{\,} - \varepsilon_{m_{i+l}}^{\,}} \bigg] \nonumber \\
& + (*) 
\sum_{K_{i+1} = 1}^{K_i-1} 
\sum_{|\tilde{n} \rangle \in D(n)}
\begin{pmatrix}
K_i \\
K_{i+1}
\end{pmatrix}
\frac{\mathcal{H}_{m_i,\tilde{n}}^{K_i - K_{i+1}}}{\varepsilon_n^{\,} - \varepsilon_{m_i}^{\,}} \, \mathcal{R}_{\tilde{n},n}^{K_{i+1}} \nonumber \\
& + \sum_{K_{i+1} = 1}^{K_i-1} \sum_{K_{i+2} = 1}^{K_{i+1}-1} 
\sum_{|m_{i+1} \rangle \neq |n \rangle}
 \sum_{|\tilde{n} \rangle \in D(n)}
\begin{pmatrix}
K_i \\
K_{i+1}
\end{pmatrix}
\begin{pmatrix}
K_{i+1} \\
K_{i+2}
\end{pmatrix}
\frac{\widetilde{\mathcal{H}}_{m_i, m_{i+1}}^{K_i - K_{i+1}}}{\varepsilon_n^{\,} - \varepsilon_{m_i}^{\,}}
\,
\frac{\mathcal{H}_{m_{i+1},\tilde{n}}^{K_{i+1} - K_{i+2}}}{\varepsilon_n^{\,} - \varepsilon_{m_{i+1}}^{\,}} \, \mathcal{R}_{\tilde{n},n}^{K_{i+2}} \nonumber \\
& + \cdots \nonumber \\
& + \cdots \nonumber \\
& + \sum_{K_{i+1} = 1}^{K_i - 1} \cdots \sum_{K_{i+l} = 1}^{K_{i+l-1} - 1} \sum_{|m_{i+1} \rangle \neq |n \rangle} \cdots \sum_{|m_{i+l-1} \rangle \neq |n \rangle} \sum_{|\tilde{n} \rangle \in D(n)} \bigg[ \begin{pmatrix}
K_i \\
K_{i+1}
\end{pmatrix} \cdots 
\begin{pmatrix}
K_{i+l-1} \\
K_{i+l}
\end{pmatrix} \nonumber \\
& \times \frac{\widetilde{\mathcal{H}}_{m_i, m_{i+1}}^{K_i - K_{i+1}}}{\varepsilon_n^{\,} - \varepsilon_{m_i}^{\,}} \times \frac{\widetilde{\mathcal{H}}_{m_{i+1}, m_{i+2}}^{K_{i+1} - K_{i+2}}}{\varepsilon_n^{\,} - \varepsilon_{m_{i+1}}^{\,}} \times \cdots \times \frac{\widetilde{\mathcal{H}}_{m_{i+l-2}, m_{i+l-1}}^{K_{i+l-2} - K_{i+l-1}}}{\varepsilon_n^{\,} - \varepsilon_{m_{i+l-2}}^{\,}} \frac{\mathcal{H}_{m_{i+l-1},\tilde{n}}^{K_{i+l-1} - K_{i+l}}}{\varepsilon_n^{\,} - \varepsilon_{m_{i+l-1}}^{\,}} \bigg] \mathcal{R}_{\tilde{n},n}^{K_{i+l}},
\end{align}
where the sequence of terms ends with maximum $l$ equalling $K_i - 1$. 

We now proceed to compute the $N^{\text{th}}$ derivative of band $n$ by taking inner product of Eq~\ref{eq:nth_deriv_eig_eqn} with $|n\rangle$, rearranging terms, introducing resolution of identity and separating out the projector to the degenerate subspace to obtain
\begin{subequations}
\begin{align}
\partial^{N} \varepsilon_{n}^{\,} = & \langle n | \partial^N H | n \rangle + \sum_{K_1=1}^{N-1} \begin{pmatrix}
N \\
K_1
\end{pmatrix}
 \langle n | \partial^{N - K_1} H | \partial^{K_1} n \rangle - \sum_{K_1=1}^{N-1} \begin{pmatrix}
N \\
K_1
\end{pmatrix}
 \partial^{N - K_1} \varepsilon_n^{\,} \langle n | \partial^{K_1} n \rangle  \\
 = & \langle n | \partial^N H | n \rangle + \sum_{K_1=1}^{N-1} 
\sum_{|m_1 \rangle \notin D(n)} 
 \begin{pmatrix}
N \\
K_1
\end{pmatrix}
 \langle n | \partial^{N - K_1} H | m_1^{\,} \rangle \langle m_1^{\,} | \partial^{K_1} n \rangle
  \nonumber \\
& 
- \sum_{K_1=1}^{N-1} 
\begin{pmatrix}
N \\
K_1
\end{pmatrix}
 \partial^{N - K_1} \varepsilon_n^{\,} \langle n | \partial^{K_1} n \rangle  
  + \sum_{K_1=1}^{N-1} 
\sum_{|\tilde{n} \rangle \in D(n)}  
  \begin{pmatrix}
N \\
K_1
\end{pmatrix}
 \langle n | \partial^{N-K_1} H | \tilde{n} \rangle \langle \tilde{n} | \partial^{K_1} n \rangle .
\end{align}
\end{subequations}
In the symbolic notation, the above reads
\begin{align}
\partial^{N} \varepsilon_{n}^{\,} = & \mathcal{H}_{n,n}^{N} + \sum_{K_1=1}^{N-1} 
\sum_{|m_1 \rangle \notin D(n)}
\begin{pmatrix}
N \\
K_1
\end{pmatrix}
\mathcal{H}_{n,m_1}^{N- K_1} \, \mathcal{O}_{m_1,n}^{K_1} 
- \sum_{K_1=1}^{N-1} \begin{pmatrix}
N \\
K_1
\end{pmatrix}
 \partial^{N - K_1} \varepsilon_n^{\,} \, \mathcal{R}^{K_1}_{n,n} \nonumber \\
& + \sum_{K_1=1}^{N-1} 
\sum_{\tilde{n} \notin D(n)}
\begin{pmatrix}
N \\
K_1
\end{pmatrix}
 \mathcal{H}^{N-K_1}_{n,\tilde{n}} \, \mathcal{R}^{K_1}_{\tilde{n},n} 
\end{align}
We now substitute for $O_{m_1}^{K_1}$:
\begin{align}
\partial^{N} \varepsilon_{n}^{\,} = & \mathcal{H}_{n,n}^{N} + \sum_{K_1=1}^{N-1} 
\sum_{|m_1 \rangle \notin D(n)}
\begin{pmatrix}
N \\
K_1
\end{pmatrix}
 \mathcal{H}_{n,m_1}^{N - K_1} \frac{\mathcal{H}_{m_1,n}^{K_1}}{\varepsilon_n^{\,} - \varepsilon_{m_1}^{\,}} \nonumber \\
& + \sum_{K_1=1}^{N-1} \sum_{K_{2} = 1}^{K_1-1} 
\sum_{|m_1 \rangle \neq |n \rangle}
\sum_{|m_{2} \rangle \neq |n \rangle}
\begin{pmatrix}
N \\
K_1
\end{pmatrix}
\begin{pmatrix}
K_1 \\
K_{2}
\end{pmatrix}
 \mathcal{H}_{n,m_1}^{N - K_1}
\frac{\widetilde{\mathcal{H}}_{m_1, m_{2}}^{K_1 - K_{2}}}{\varepsilon_n^{\,} - \varepsilon_{m_1}^{\,}} \frac{\mathcal{H}_{m_{2},n}^{K_{2}}}{\varepsilon_n^{\,} - \varepsilon_{m_{2}}^{\,}} \nonumber \\
& + \cdots \nonumber \\
& + \cdots \nonumber \\
& + \sum_{K_1=1}^{N-1} \sum_{K_{2} = 1}^{K_1-1}
\cdots
\sum_{K_{N-1} = 1}^{K_{N-2} - 1}
\sum_{|m_1 \rangle \neq |n \rangle}
\sum_{|m_{2} \rangle \neq |n \rangle}
\cdots
\sum_{|m_{N-1} \rangle \neq |n \rangle}
\begin{pmatrix}
N \\
K_1
\end{pmatrix}
\begin{pmatrix}
K_1 \\
K_{2}
\end{pmatrix} 
\cdots
\begin{pmatrix}
K_{N-2} \\
K_{N-1}
\end{pmatrix} 
\bigg[
\nonumber \\
& \qquad \times \mathcal{H}_{n,m_1}^{N - K_1} \left( \frac{\widetilde{\mathcal{H}}_{m_1, m_{2}}^{K_1 - K_{2}}}{\varepsilon_n^{\,} - \varepsilon_{m_1}^{\,}} \times \frac{\widetilde{\mathcal{H}}_{m_{2}, m_{3}}^{K_{2} - K_{3}}}{\varepsilon_n^{\,} - \varepsilon_{m_{2}}^{\,}} \times \cdots \times \frac{\widetilde{\mathcal{H}}_{m_{N-2}, m_{N-1}}^{K_{N-2} - K_{N-1}}}{\varepsilon_n^{\,} - \varepsilon_{m_{N-2}}^{\,}} \right) 
\frac{\mathcal{H}_{m_{N-1},n}^{K_{N-1}}}{\varepsilon_n^{\,} - \varepsilon_{m_{N-1}}^{\,}} \bigg]
\nonumber \\
& - (*1) \sum_{K_1=1}^{N-1} \begin{pmatrix}
N \\
K_1
\end{pmatrix}
 \partial^{N - K_1} \varepsilon_n^{\,} \, \mathcal{R}^{K_1}_{n,n} \nonumber \\
& + (*2) \sum_{K_1=1}^{N-1} 
\sum_{|\tilde{n} \rangle \in D(n)}
\begin{pmatrix}
N \\
K_1
\end{pmatrix}
 \mathcal{H}^{N-K_1}_{n,\tilde{n}} \, \mathcal{R}^{K_1}_{\tilde{n},n} \nonumber \\
 & + \sum_{K_1=1}^{N-1}  
\sum_{K_{2} = 1}^{K_1-1} 
\sum_{|m_1 \rangle \notin D(n)}
\sum_{|\tilde{n} \rangle \in D(n)}
\begin{pmatrix}
N \\
K_1
\end{pmatrix}
\begin{pmatrix}
K_1 \\
K_{2}
\end{pmatrix}
 \mathcal{H}_{n,m_1}^{N - K_1} 
\frac{\mathcal{H}_{m_1,\tilde{n}}^{K_1 - K_{2}}}{\varepsilon_n^{\,} - \varepsilon_{m_1}^{\,}} \, \mathcal{R}^{K_2}_{\tilde{n},n}
 \nonumber \\
 & + \cdots \nonumber \\
 & + \cdots \nonumber \\
 & + \sum_{K_1=1}^{N-1}  
\sum_{K_{2} = 1}^{K_1-1}
\cdots
\sum_{K_{N-1} = 1}^{K_{N-2} - 1}
\sum_{|m_1 \rangle \notin D(n)}
\cdots
\sum_{|m_{N-2} \rangle \notin D(n)}
\sum_{|\tilde{n} \rangle \in D(n)}
\begin{pmatrix}
N \\
K_1
\end{pmatrix}
\begin{pmatrix}
K_1 \\
K_{2}
\end{pmatrix} 
\begin{pmatrix}
K_2 \\
K_{3}
\end{pmatrix} 
\cdots
\begin{pmatrix}
K_{N-2} \\
K_{N-1}
\end{pmatrix}
\bigg[
\nonumber \\
& \qquad \times \mathcal{H}_{n,m_1}^{N - K_1} \left( \frac{\widetilde{\mathcal{H}}_{m_1, m_{2}}^{K_1 - K_{2}}}{\varepsilon_n^{\,} - \varepsilon_{m_1}^{\,}} \times \frac{\widetilde{\mathcal{H}}_{m_{2}, m_{3}}^{K_{2} - K_{3}}}{\varepsilon_n^{\,} - \varepsilon_{m_{2}}^{\,}} \times \cdots \times \frac{\widetilde{\mathcal{H}}_{m_{N-3}, m_{N-2}}^{K_{N-3} - K_{N-2}}}{\varepsilon_n^{\,} - \varepsilon_{m_{N-3}}^{\,}} \right) 
\frac{\mathcal{H}_{m_{N-2},\tilde{n}}^{K_{N-2} - K_{N-1}}}{\varepsilon_n^{\,} - \varepsilon_{m_{N-2}}^{\,}} \bigg] \mathcal{R}^{K_{N-1}}_{\tilde{n},n} .
\label{eq:band_deriv_with_residue}
\end{align}
\textbf{Theorem}: The term $(*1)$ cancels with terms from $(2*)$ onwards so that the band derivative takes the explicitly calculable form:
\begin{align}
\partial^{N} \varepsilon_{n}^{\,} = & \mathcal{H}_{n,n}^{N} + \sum_{K_1=1}^{N-1} \begin{pmatrix}
N \\
K_1
\end{pmatrix}
\sum_{|m_1 \rangle \notin D(n)}
 \mathcal{H}_{n,m_1}^{N - K_1} \frac{\mathcal{H}_{m_1,n}^{K_1}}{\varepsilon_n^{\,} - \varepsilon_{m_1}^{\,}} \nonumber \\
& + \sum_{K_1=1}^{N-1} \sum_{K_{2} = 1}^{K_1-1} 
\sum_{|m_1 \rangle \notin D(n)}
\sum_{|m_{2} \rangle \notin D(n)}
\begin{pmatrix}
N \\
K_1
\end{pmatrix}
\begin{pmatrix}
K_1 \\
K_{2}
\end{pmatrix}
 \mathcal{H}_{n,m_1}^{N - K_1}
\frac{\widetilde{\mathcal{H}}_{m_1, m_{2}}^{K_1 - K_{2}}}{\varepsilon_n^{\,} - \varepsilon_{m_1}^{\,}} \frac{\mathcal{H}_{m_{2},n}^{K_{2}}}{\varepsilon_n^{\,} - \varepsilon_{m_{2}}^{\,}} \nonumber \\
& + \cdots \nonumber \\
& + \cdots \nonumber \\
& + \sum_{K_1=1}^{N-1} \sum_{K_{2} = 1}^{K_1-1}
\cdots
\sum_{K_{N-1} = 1}^{K_{N-2} - 1}
\sum_{|m_1 \rangle \notin D(n)}
\sum_{|m_{2} \rangle \notin D(n)}
\cdots
\sum_{|m_{N-1} \rangle \notin D(n)}
\begin{pmatrix}
N \\
K_1
\end{pmatrix}
\begin{pmatrix}
K_1 \\
K_{2}
\end{pmatrix} 
\begin{pmatrix}
K_2 \\
K_{3}
\end{pmatrix} 
\cdots
\begin{pmatrix}
K_{N-2} \\
K_{N-1}
\end{pmatrix} 
\bigg[
\nonumber \\
& \qquad \times \mathcal{H}_{n,m_1}^{N - K_1} \left( \frac{\widetilde{\mathcal{H}}_{m_1, m_{2}}^{K_1 - K_{2}}}{\varepsilon_n^{\,} - \varepsilon_{m_1}^{\,}} \times \frac{\widetilde{\mathcal{H}}_{m_{2}, m_{3}}^{K_{2} - K_{3}}}{\varepsilon_n^{\,} - \varepsilon_{m_{2}}^{\,}} \times \cdots \times \frac{\widetilde{\mathcal{H}}_{m_{N-2}, m_{N-1}}^{K_{N-2} - K_{N-1}}}{\varepsilon_n^{\,} - \varepsilon_{m_{N-2}}^{\,}} \right) 
 \frac{\mathcal{H}_{m_{N-1},n}^{K_{N-1}}}{\varepsilon_n^{\,} - \varepsilon_{m_{N-1}}^{\,}} \bigg] .
\label{eq:band_deriv_final}
\end{align}
Furthermore, for any $|\tilde{n} \rangle$ distinct from, but degenerate to $| n \rangle$, the following is satisfied
\begin{align}
0 = & \mathcal{H}_{\tilde{n},n}^{N} + \sum_{K_1=1}^{N-1} \begin{pmatrix}
N \\
K_1
\end{pmatrix}
\sum_{|m_1 \rangle \notin D(n)}
 \mathcal{H}_{\tilde{n},m_1}^{N - K_1} \frac{\mathcal{H}_{m_1,n}^{K_1}}{\varepsilon_n^{\,} - \varepsilon_{m_1}^{\,}} \nonumber \\
& + \sum_{K_1=1}^{N-1} \sum_{K_{2} = 1}^{K_1-1} 
\sum_{|m_1 \rangle \notin D(n)}
\sum_{|m_{2} \rangle \notin D(n)}
\begin{pmatrix}
N \\
K_1
\end{pmatrix}
\begin{pmatrix}
K_1 \\
K_{2}
\end{pmatrix}
 \mathcal{H}_{\tilde{n},m_1}^{N - K_1}
\frac{\widetilde{\mathcal{H}}_{m_1, m_{2}}^{K_1 - K_{2}}}{\varepsilon_n^{\,} - \varepsilon_{m_1}^{\,}} \frac{\mathcal{H}_{m_{2},n}^{K_{2}}}{\varepsilon_n^{\,} - \varepsilon_{m_{2}}^{\,}} \nonumber \\
& + \cdots \nonumber \\
& + \cdots \nonumber \\
& + \sum_{K_1=1}^{N-1} \sum_{K_{2} = 1}^{K_1-1}
\cdots
\sum_{K_{N-1} = 1}^{K_{N-2} - 1}
\sum_{|m_1 \rangle \notin D(n)}
\sum_{|m_{2} \rangle \notin D(n)}
\cdots
\sum_{|m_{N-1} \rangle \notin D(n)}
\begin{pmatrix}
N \\
K_1
\end{pmatrix}
\begin{pmatrix}
K_1 \\
K_{2}
\end{pmatrix} 
\begin{pmatrix}
K_2 \\
K_{3}
\end{pmatrix} 
\cdots
\begin{pmatrix}
K_{N-2} \\
K_{N-1}
\end{pmatrix} 
\bigg[
\nonumber \\
& \qquad \times \mathcal{H}_{\tilde{n},m_1}^{N - K_1} \left( \frac{\widetilde{\mathcal{H}}_{m_1, m_{2}}^{K_1 - K_{2}}}{\varepsilon_n^{\,} - \varepsilon_{m_1}^{\,}} \times \frac{\widetilde{\mathcal{H}}_{m_{2}, m_{3}}^{K_{2} - K_{3}}}{\varepsilon_n^{\,} - \varepsilon_{m_{2}}^{\,}} \times \cdots \times \frac{\widetilde{\mathcal{H}}_{m_{N-2}, m_{N-1}}^{K_{N-2} - K_{N-1}}}{\varepsilon_n^{\,} - \varepsilon_{m_{N-2}}^{\,}} \right) 
 \frac{\mathcal{H}_{m_{N-1},n}^{K_{N-1}}}{\varepsilon_n^{\,} - \varepsilon_{m_{N-1}}^{\,}} \bigg] .
\label{eq:band_deriv_degen}
\end{align}

\textbf{Proof}: It is easy to see that this form for $\partial^N \varepsilon_n^{\,}$ is satisfied for $N = 1, 2$ and $3$, by explicit computation and comparison with Eq~\ref{eq:Taylor_explicit}. As for Eq~\ref{eq:band_deriv_degen}, its RHS has the same form as that of $\partial^N \varepsilon_n^{\,}$, with the difference that the leftmost bra is $\langle \tilde{n}|$ instead of $\langle n |$. As Eqs~\ref{eq:overlap_1st_deriv_degen} and~\ref{eq:second_deriv_overlap_nn} show, this too holds for $N = 1, 2$. For higher values of $N$, we prove both the results by the principle of mathematical induction. Let us begin with $\partial^N \varepsilon_n^{\,}$. Assume that the expressions above are satisfied for $\partial^{M} \varepsilon_n^{\,}$ for $1 \leqslant M \leqslant N-1$. To prove that they are satisfied for $M = N$, we have to show that the terms from $(2*)$ cancel with $(1*)$ in Eq~\ref{eq:band_deriv_with_residue}. Now $(2*)$ is a sum of several individual terms which take the form
\begin{multline}
 \begin{pmatrix}
N \\
K_1
\end{pmatrix}
\begin{pmatrix}
K_1 \\
K_{2}
\end{pmatrix} 
\begin{pmatrix}
K_2 \\
K_{3}
\end{pmatrix} 
\cdots
\begin{pmatrix}
K_{l-1} \\
K_{l}
\end{pmatrix} \\
\times
\mathcal{H}_{n,m_1}^{N - K_1} \left( \frac{\widetilde{\mathcal{H}}_{m_1, m_{2}}^{K_1 - K_{2}}}{\varepsilon_n^{\,} - \varepsilon_{m_1}^{\,}} \times \frac{\widetilde{\mathcal{H}}_{m_{2}, m_{3}}^{K_{2} - K_{3}}}{\varepsilon_n^{\,} - \varepsilon_{m_{2}}^{\,}} \times \cdots \times \frac{\widetilde{\mathcal{H}}_{m_{l-2}, m_{l-1}}^{K_{l-2} - K_{l-1}}}{\varepsilon_n^{\,} - \varepsilon_{m_{l-2}}^{\,}} \right) 
 \frac{\mathcal{H}_{m_{l-1},\tilde{n}}^{K_{l-1} - K_{l}}}{\varepsilon_n^{\,} - \varepsilon_{m_{l-1}}^{\,}} \bigg] \mathcal{R}^{K_{l}}_{\tilde{n},n},
\end{multline}
for \emph{all possible} ordered sequences of $l$ natural numbers $\{K_1, K_2, \cdots, K_l \}$ such that $N > K_1 > K_2 > \cdots > K_l \geqslant 1$ and $1 \leqslant l \leqslant N-1$. The fact that every such \emph{ordered} sequence occurs in the sum, and that too only once is ensured by the cascaded $K_i$ sums in Eq~\ref{eq:band_deriv_with_residue}
\begin{equation}
\sum_{K_1=1}^{N-1}  
\sum_{K_{2} = 1}^{K_1-1}
\cdots
\sum_{K_{l} = 1}^{K_{l-1} - 1}
\end{equation}
Let us denote by $A_l$, the set of all such sequences of length $l$ (with $1 \leqslant l \leqslant N-1$). That is, $A_l$ = Set of all ordered sequences $ \{K_1, K_2, \cdots, K_l \} $ such that $K_i \in \mathbb{N}$ and $N > K_1 > K_2 > \cdots > K_l \geqslant 1$. It is then easy to see that
\begin{equation}
\sum_{K_1=1}^{N-1}  
\sum_{K_{2} = 1}^{K_1-1}
\cdots
\sum_{K_{l} = 1}^{K_{l-1} - 1}
\equiv
\sum_{\{K_1, \cdots, K_l\} \in A_l}.
\label{eq:sum_over_A_l}
\end{equation}

Let us define another set $B_l$ which is the set of all ordered sequences of $l$ natural numbers $\{ K_l, \widetilde{K}_1, \widetilde{K}_2, \cdots \widetilde{K}_{l-1} \}$ such that $N - l \geqslant K_l \geqslant 1$ and $N - K_l > \widetilde{K}_1 > \widetilde{K}_2 > \cdots > \widetilde{K}_{l-1} \geqslant 1$. It is also easy to see that
\begin{equation}
\sum_{K_l=1}^{N-l}  
\sum_{\widetilde{K}_{1} = 1}^{N-K_1-1}
\sum_{\widetilde{K}_{2} = 1}^{\widetilde{K}_1-1}
\cdots
\sum_{\widetilde{K}_{l-1} = 1}^{\widetilde{K}_{l-2} - 1}
\equiv
\sum_{\{K_l, \widetilde{K}_1, \cdots \widetilde{K}_{l-1}\} \in B_l}.
\label{eq:sum_over_B_l}
\end{equation}
\textbf{Lemma}: There exists a bijection between $A_l$ and $B_l$ for \emph{each} $l$. \\
\textbf{Proof}: Let us define $F: A_l \rightarrow B_l$, with $F(K_1, K_2, \cdots, K_{l-1}, K_l) = \{K_l,  K_1 - K_l, K_2 - K_l, \cdots, K_{l-1} - K_l \}$. Since $N > K_1 > K_2 > \cdots > K_l \geqslant 1$, it is easy to see that the range is indeed contained in $B_l$ since $N - l \geqslant K_l \geqslant 1$ and $N - K_l > K_1 - K_l > K_2 - K_l > \cdots > K_{l-1} - K_l > 0$. (The last inequality $K_{l-1} - K_l > 0$ is equivalent to $K_{l-1} - K_l \geqslant 1$ since we are dealing with natural numbers). \\

Let us define the function $G: B_l \rightarrow A_l$ with $G(K_l, \widetilde{K}_1, \widetilde{K}_2, \cdots \widetilde{K}_{l-1}) = \{\widetilde{K}_1 + K_l, \widetilde{K}_2 + K_l, \cdots ,\widetilde{K}_{l-1} + K_l, K_l \}$. Since $N - l \geqslant K_l \geqslant 1$ and $N - K_l > \widetilde{K}_1 > \widetilde{K}_2 > \cdots > \widetilde{K}_{l-1} \geqslant 1$, it is easy to see that $N > \widetilde{K}_1 + K_l > \widetilde{K}_2 + K_l > \cdots > \widetilde{K}_{l-1} + K_l > K_l \geqslant 1$ so that the range of this function is contained in $A_l$. Since $F$ and $G$ are clearly inverses of each other, they are both bijections. In particular, this means that iterating over each member of $A_l$ and evaluating a function $\psi_l^{\,}$ defined on $A_l$ is equivalent to iterating over each member of $B_l$:
\begin{align}
& \sum_{l=1}^{N-1} \sum_{\{K_1, \cdots, K_l\} \in A_l} \psi_l^{\,}(K_1, \cdots, K_l) \equiv \sum_{l=1}^{N-1} \sum_{\{K_l, \widetilde{K}_1, \cdots, \widetilde{K}_{l-1}\} \in B_l } \psi_l^{\,} (G(K_l, \widetilde{K}_1, \cdots, \widetilde{K}_{l-1})) \nonumber \\
\implies & \sum_{l=1}^{N-1} \sum_{K_1=1}^{N-1}  
\sum_{K_{2} = 1}^{K_1-1}
\cdots
\sum_{K_{l} = 1}^{K_{l-1} - 1}
\psi_l^{\,}(K_1, K_2, \cdots, K_l) \nonumber \\
& = \sum_{l=1}^{N-1}
\sum_{K_l=1}^{N-l}  
\sum_{\widetilde{K}_{1} = 1}^{N - K_1 - 1}
\sum_{\widetilde{K}_{2} = 1}^{\widetilde{K}_1 - 1}
\cdots
\sum_{\widetilde{K}_{l-1} = 1}^{\widetilde{K}_{l-2} - 1}
\psi_l^{\,}(G(K_l, \widetilde{K}_1, \widetilde{K}_2, \cdots , \widetilde{K}_{l-1})).
\end{align}
The sums on LHS and RHS exhaustively go through all the elements in $A_l$ and $B_l$ respectively, as seen from Eqs~\ref{eq:sum_over_A_l} and \ref{eq:sum_over_B_l}. The function $g$ is needed on the RHS since, for each element of $B_l$, we need to feed its unique counterpart in $A_l$ as an input to the function $\psi_l^{\,}$, to sum over.  \\

\begin{figure}[t]
\centering
\includegraphics{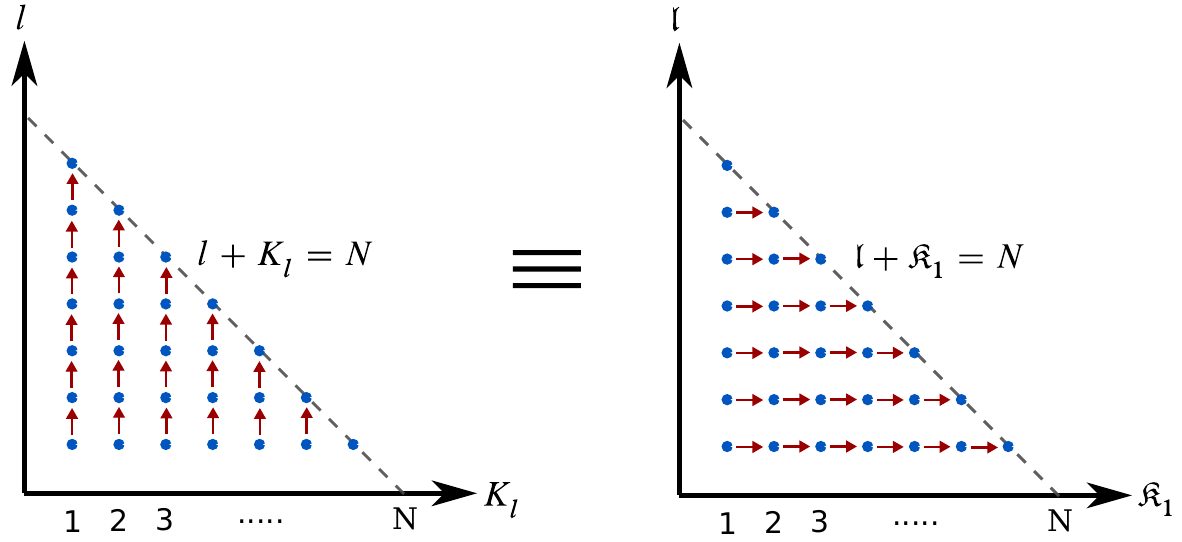}
\caption{Two equivalent ways of iterating over a grid of points. The left panel represents the sum $\sum_{l=1}^{N-1} \sum_{K_l=1}^{N-l}$, while the right panel denotes $\sum_{\mathfrak{K}_1=1}^{N-1} \sum_{\mathfrak{l}=1}^{N-\mathfrak{K}_1}$.}
\label{fig:Proof}
\end{figure}

To be able to interchange the $K_l$ and $l$ sums in the RHS, let us define $C$ to be the set of all ordered pairs $\{l, K_l \}$ with $1 \leqslant l \leqslant N -1$ and $1 \leqslant K_l \leqslant N - l$ and $D$ to be the set of all ordered pairs $\{ \mathfrak{K}_1, \mathfrak{l} \}$ with $1 \leqslant \mathfrak{K}_1 \leqslant N - 1$ and $1 \leqslant \mathfrak{l} \leqslant N - \mathfrak{K}_1$. As before, we define functions $\mathcal{F}: C \rightarrow D$ with $\mathcal{F} (l, K_l) = \{ K_l, l \}$ and $\mathcal{G}: D \rightarrow C$ with $ \mathcal{G} (\mathfrak{K}_1, \mathfrak{l}) = \{ \mathfrak{l}, \mathfrak{K}_1 \}$. It is easy to check that $\text{Range} [\mathcal{F}] \subseteq D$, $\text{Range} [\mathcal{G}] \subseteq C$, and the functions are inverses of each other, allowing us to interchange the original sums:
\begin{equation}
\sum_{l=1}^{N-1}
\sum_{K_l=1}^{N-l} \equiv 
\sum_{\mathfrak{K}_1=1}^{N-1}
\sum_{\mathfrak{l}=1}^{N-\mathfrak{K}_1}.
\label{eq:equiv_sum}
\end{equation}
This is geometrically depicted in Fig~\ref{fig:Proof}, where the grid points in the left panel are being summed over exhaustively by the sum on LHS in Eq~\ref{eq:equiv_sum}. Clearly the sum on RHS is just a different but equivalent approach to summing over the \emph{same} set of points, depicted in the right panel.
Thus, we have
\begin{align}
&\sum_{l=1}^{N-1} \sum_{K_1=1}^{N-1}  
\sum_{K_{2} = 1}^{K_1-1}
\cdots
\sum_{K_{l} = 1}^{K_{l-1} - 1}
\psi_l^{\,}(K_1, K_2, \cdots, K_l) \nonumber \\
& = \sum_{\mathfrak{K}_1=1}^{N-1}
\sum_{\mathfrak{l}=1}^{N-\mathfrak{K}_1}
\sum_{\widetilde{K}_{1} = 1}^{N - \mathfrak{K}_1 - 1}
\sum_{\widetilde{K}_{2} = 1}^{\widetilde{K}_1 - 1}
\cdots
\sum_{\widetilde{K}_{\mathfrak{l}-1} = 1}^{\widetilde{K}_{\mathfrak{l}-2} - 1}
\psi_\mathfrak{l}^{\,}(G(\mathfrak{K}_1, \widetilde{K}_1, \widetilde{K}_2, \cdots , \widetilde{K}_{\mathfrak{l}-1}))
\nonumber \\
& = \sum_{\mathfrak{K}_1=1}^{N-1}
\sum_{\mathfrak{l}=1}^{N-\mathfrak{K}_1}
\sum_{\widetilde{K}_{1} = 1}^{N - \mathfrak{K}_1 - 1}
\sum_{\widetilde{K}_{2} = 1}^{\widetilde{K}_1 - 1}
\cdots
\sum_{\widetilde{K}_{\mathfrak{l}-1} = 1}^{\widetilde{K}_{\mathfrak{l}-2} - 1}
\psi_\mathfrak{l}^{\,}(\widetilde{K}_1 + \mathfrak{K}_1, \widetilde{K}_2 + \mathfrak{K}_1, \cdots , \widetilde{K}_{\mathfrak{l}-1} + \mathfrak{K}_1, \mathfrak{K}_1).
\end{align}
For convenience, we shall henceforth refer to $\mathfrak{K}_1$ as $K_1$ and $\mathfrak{l}$ as $l$. Applying the above results to $(2*)$ in Eq~\ref{eq:band_deriv_with_residue} (suppressing the $|m_i \rangle \notin D(n)$ and $|\tilde{n} \rangle \in D(n)$ sums for the sake convenience), we get 
\begin{subequations}
\begin{align}
& (2*) \nonumber \\
&= \sum_{l=1}^{N-1} \sum_{K_1=1}^{N-1}  
\sum_{K_{2} = 1}^{K_1-1}
\cdots
\sum_{K_{l} = 1}^{K_{l-1} - 1} \bigg[
\begin{pmatrix}
N \\
K_1
\end{pmatrix}
\begin{pmatrix}
K_1 \\
K_{2}
\end{pmatrix} 
\begin{pmatrix}
K_2 \\
K_{3}
\end{pmatrix} 
\cdots
\begin{pmatrix}
K_{l-1} \\
K_{l}
\end{pmatrix} \\
& \qquad \times
\mathcal{H}_{n,m_1}^{N - K_1} \left( \frac{\widetilde{\mathcal{H}}_{m_1, m_{2}}^{K_1 - K_{2}}}{\varepsilon_n^{\,} - \varepsilon_{m_1}^{\,}} \times \frac{\widetilde{\mathcal{H}}_{m_{2}, m_{3}}^{K_{2} - K_{3}}}{\varepsilon_n^{\,} - \varepsilon_{m_{2}}^{\,}} \times \cdots \times \frac{\widetilde{\mathcal{H}}_{m_{l-2}, m_{l-1}}^{K_{l-2} - K_{l-1}}}{\varepsilon_n^{\,} - \varepsilon_{m_{l-2}}^{\,}} \right) 
 \frac{\mathcal{H}_{m_{l-1},\tilde{n}}^{K_{l-1} - K_{l}}}{\varepsilon_n^{\,} - \varepsilon_{m_{l-1}}^{\,}} \bigg] \mathcal{R}^{K_{l}}_{\tilde{n},n}
\nonumber \\
& = \sum_{K_1=1}^{N-1}
\sum_{l=1}^{N-K_1}
\sum_{\widetilde{K}_{1} = 1}^{N - K_1 - 1}
\sum_{\widetilde{K}_{2} = 1}^{\widetilde{K}_1 - 1}
\cdots
\sum_{\widetilde{K}_{l-1} = 1}^{\widetilde{K}_{l-2} - 1}  \bigg[ \nonumber \\
& \qquad \begin{pmatrix}
N \\
\widetilde{K}_1 + K_1
\end{pmatrix}
\begin{pmatrix}
\widetilde{K}_1 + K_1 \\
\widetilde{K}_{2} + K_1
\end{pmatrix} 
\begin{pmatrix}
\widetilde{K}_2 + K_1 \\
\widetilde{K}_{3} + K_1
\end{pmatrix} 
\cdots
\begin{pmatrix}
\widetilde{K}_{l-2} + K_1 \\
\widetilde{K}_{l-1} + K_1
\end{pmatrix}
\begin{pmatrix}
\widetilde{K}_{l-1} + K_1 \\
K_1
\end{pmatrix} \nonumber \\
& \qquad \times
\mathcal{H}_{n,m_1}^{N - K_1 - \widetilde{K}_1} \left( \frac{\widetilde{\mathcal{H}}_{m_1, m_{2}}^{\widetilde{K}_1 - \widetilde{K}_{2}}}{\varepsilon_n^{\,} - \varepsilon_{m_1}^{\,}} \times \frac{\widetilde{\mathcal{H}}_{m_{2}, m_{3}}^{\widetilde{K}_{2} - \widetilde{K}_{3}}}{\varepsilon_n^{\,} - \varepsilon_{m_{2}}^{\,}} \times \cdots \times \frac{\widetilde{\mathcal{H}}_{m_{l-2}, m_{l-1}}^{\widetilde{K}_{l-2} - \widetilde{K}_{l-1}}}{\varepsilon_n^{\,} - \varepsilon_{m_{l-2}}^{\,}} \right) 
 \frac{\mathcal{H}_{m_{l-1},\tilde{n}}^{\widetilde{K}_{l-1}}}{\varepsilon_n^{\,} - \varepsilon_{m_{l-1}}^{\,}} \bigg] \mathcal{R}^{K_{1}}_{\tilde{n},n} .
\end{align}
\end{subequations}
Now for the last ingredient of the proof, we use the equivalent expression of the product of the combinatorial factors:
\begin{align}
\begin{pmatrix}
N \\
\widetilde{K}_1 + K_1
\end{pmatrix}
&
\begin{pmatrix}
\widetilde{K}_1 + K_1 \\
\widetilde{K}_{2} + K_1
\end{pmatrix} 
\begin{pmatrix}
\widetilde{K}_2 + K_1 \\
\widetilde{K}_{3} + K_1
\end{pmatrix} 
\cdots
\begin{pmatrix}
\widetilde{K}_{l-2} + K_1 \\
\widetilde{K}_{l-1} + K_1
\end{pmatrix}
\begin{pmatrix}
\widetilde{K}_{l-1} + K_1 \\
K_1
\end{pmatrix}
\nonumber \\
& = 
\begin{pmatrix}
N \\
K_1
\end{pmatrix}
\begin{pmatrix}
N - K_1 \\
\widetilde{K}_{1}
\end{pmatrix} 
\begin{pmatrix}
\widetilde{K}_1 \\
\widetilde{K}_{2}
\end{pmatrix} 
\begin{pmatrix}
\widetilde{K}_2 \\
\widetilde{K}_{3}
\end{pmatrix} 
\cdots
\begin{pmatrix}
\widetilde{K}_{l-2} \\
\widetilde{K}_{l-1}
\end{pmatrix}.
\end{align}
Collecting all of the results above, we obtain for $(2*)$
\begin{align}
(2*) 
& = \sum_{K_1=1}^{N-1} \,
\sum_{l=1}^{N-K_1} \,
\sum_{\widetilde{K}_{1} = 1}^{N - K_1 - 1} \,
\sum_{\widetilde{K}_{2} = 1}^{\widetilde{K}_1 - 1}
\cdots
\sum_{\widetilde{K}_{l-1} = 1}^{\widetilde{K}_{l-2} - 1}   \bigg[ 
\begin{pmatrix}
N \\
K_1
\end{pmatrix}
\begin{pmatrix}
N - K_1 \\
\widetilde{K}_{1}
\end{pmatrix} 
\begin{pmatrix}
\widetilde{K}_1 \\
\widetilde{K}_{2}
\end{pmatrix} 
\begin{pmatrix}
\widetilde{K}_2 \\
\widetilde{K}_{3}
\end{pmatrix} 
\cdots
\begin{pmatrix}
\widetilde{K}_{l-2} \\
\widetilde{K}_{l-1}
\end{pmatrix}
\nonumber \\
& \qquad \times
\mathcal{H}_{n,m_1}^{N - K_1 - \widetilde{K}_1} \left( \frac{\widetilde{\mathcal{H}}_{m_1, m_{2}}^{\widetilde{K}_1 - \widetilde{K}_{2}}}{\varepsilon_n^{\,} - \varepsilon_{m_1}^{\,}} \times \frac{\widetilde{\mathcal{H}}_{m_{2}, m_{3}}^{\widetilde{K}_{2} - \widetilde{K}_{3}}}{\varepsilon_n^{\,} - \varepsilon_{m_{2}}^{\,}} \times \cdots \times \frac{\widetilde{\mathcal{H}}_{m_{l-2}, m_{l-1}}^{\widetilde{K}_{l-2} - \widetilde{K}_{l-1}}}{\varepsilon_n^{\,} - \varepsilon_{m_{l-2}}^{\,}} \right) 
 \frac{\mathcal{H}_{m_{l-1},\tilde{n}}^{\widetilde{K}_{l-1}}}{\varepsilon_n^{\,} - \varepsilon_{m_{l-1}}^{\,}} \bigg] \mathcal{R}^{K_{1}}_{\tilde{n},n} 
 \nonumber \\
 & = \sum_{K_1=1}^{N-1} \begin{pmatrix}
N \\
K_1
\end{pmatrix}
\mathcal{R}^{K_{1}}_{\tilde{n},n} \bigg[ \mathcal{H}_{n, \tilde{n}}^{N-K_1} + \sum_{\widetilde{K}_1=1}^{N - K_1 -1} \begin{pmatrix}
N - K_1 \\
\widetilde{K}_1
\end{pmatrix}
\sum_{|m_1 \rangle \notin D(n)}
 \mathcal{H}_{n,m_1}^{N - K_1 - \widetilde{K}_1} \frac{\mathcal{H}_{m_1, \tilde{n}}^{\widetilde{K}_1}}{\varepsilon_n^{\,} - \varepsilon_{m_1}^{\,}} \nonumber \\
 & \qquad + \sum_{\widetilde{K}_1=1}^{N - K_1-1} \, \sum_{\widetilde{K}_{2} = 1}^{\widetilde{K}_1-1} \,
\sum_{|m_1 \rangle \notin D(n)} \,
\sum_{|m_{2} \rangle \notin D(n)}
\begin{pmatrix}
N - K_1 \\
\widetilde{K}_1
\end{pmatrix}
\begin{pmatrix}
\widetilde{K}_1 \\
\widetilde{K}_{2}
\end{pmatrix}
 \mathcal{H}_{n,m_1}^{N - K_1 - \widetilde{K}_1}
\frac{\widetilde{\mathcal{H}}_{m_1, m_{2}}^{\widetilde{K}_1 - \widetilde{K}_{2}}}{\varepsilon_n^{\,} - \varepsilon_{m_1}^{\,}} \frac{\mathcal{H}_{m_{2},\tilde{n}}^{\widetilde{K}_{2}}}{\varepsilon_n^{\,} - \varepsilon_{m_{2}}^{\,}} \nonumber \\
& \qquad + \cdots \nonumber \\
& \qquad + \cdots \nonumber \\
& \qquad + \sum_{\widetilde{K}_1=1}^{N - K_1 -1} \, \sum_{\widetilde{K}_{2} = 1}^{\widetilde{K}_1-1}
\cdots
\sum_{\widetilde{K}_{N - K_1 -1} = 1}^{\widetilde{K}_{N - K_1 -2} - 1} \,
\sum_{|m_1 \rangle \notin D(n)} \,
\sum_{|m_{2} \rangle \notin D(n)}
\cdots
\sum_{|m_{N - K_1 - 1} \rangle \notin D(n)}
\bigg\{
 \nonumber \\
& \qquad \qquad \begin{pmatrix}
N - K_1 \\
\widetilde{K}_1
\end{pmatrix}
\begin{pmatrix}
\widetilde{K}_1 \\
\widetilde{K}_{2}
\end{pmatrix} 
\begin{pmatrix}
\widetilde{K}_2 \\
\widetilde{K}_{3}
\end{pmatrix} 
\cdots
\begin{pmatrix}
\widetilde{K}_{N - K_1 - 2} \\
\widetilde{K}_{N - K_1 - 1}
\end{pmatrix} 
\nonumber \\
& \qquad \qquad \times \mathcal{H}_{n,m_1}^{N - K_1 - \widetilde{K}_1} \left( \frac{\widetilde{\mathcal{H}}_{m_1, m_{2}}^{\widetilde{K}_1 - \widetilde{K}_{2}}}{\varepsilon_n^{\,} - \varepsilon_{m_1}^{\,}} \times \frac{\widetilde{\mathcal{H}}_{m_{2}, m_{3}}^{\widetilde{K}_{2} - \widetilde{K}_{3}}}{\varepsilon_n^{\,} - \varepsilon_{m_{2}}^{\,}} \times \cdots \times \frac{\widetilde{\mathcal{H}}_{m_{N - K_1 - 2}, m_{N - K_1 - 1}}^{\widetilde{K}_{N - K_1 - 2} - \widetilde{K}_{N - K_1 - 1}}}{\varepsilon_n^{\,} - \varepsilon_{m_{N - K_1 - 2}}^{\,}} \right) 
 \nonumber \\
& \qquad \qquad \times \frac{\mathcal{H}_{m_{N - K_1 - 1}, \tilde{n}}^{K_{N - K_1 - 1}}}{\varepsilon_n^{\,} - \varepsilon_{m_{N - K_1 - 1}}^{\,}} \bigg\} \bigg].
\end{align}
On inspection, we observe that the terms within the square brackets $[\,]$ precisely add up to give $\partial^{N - K_1} \varepsilon_n^{\,}$ when $|\tilde{n} \rangle = |n \rangle$ (according to Eq~\ref{eq:band_deriv_final}) and vanish when $|\tilde{n} \rangle \neq |n \rangle$ (as per Eq~\ref{eq:band_deriv_degen}). This is the case because $N - K_1 < N$ or equivalently $N - K_1 \leqslant N - 1$, and we have assumed that the form of the band derivative given in Eq~\ref{eq:band_deriv_final} holds up to $N - 1$. Thus some of the terms in $(2*)$ vanish while the rest add up to
\begin{equation}
(2*) = \sum_{K_1 = 1}^{N - 1} \begin{pmatrix}
N \\
K_1
\end{pmatrix}
\partial^{N - K_1} \varepsilon_n^{\,} \,
\mathcal{R}_{n,n}^{K_1},
\end{equation}
which exactly cancels out with $(1*)$. As for the proof of Eq~\ref{eq:band_deriv_degen} at $M = N$, we take overlap of Eq~\ref{eq:nth_deriv_eig_eqn} with $|\tilde{n} \rangle$ instead of $|n\rangle$, which causes the term containing $\partial^{N} \varepsilon_n^{\,}$ to vanish. The rest of the proof proceeds exactly the same way as for the proof for $\partial^N \varepsilon_n^{\,}$, with the extra assumption that the projection of the derivatives of the Hamiltonian $\partial^M H$ to the degenerate subspace are diagonal and proportional to identity matrix. This ensures that the derivatives for the degenerate bands are also equal. Thus we have proved that the forms given in Eq~\ref{eq:band_deriv_final} and~\ref{eq:band_deriv_degen} hold for $N$ as long as they hold up from $1$ to $N - 1$. Since they have already been shown to hold for $N = 1, 2$, it follows that they hold for all $N$.

\end{appendix}

\medskip

%
\bibliographystyle{apsrev4-2}

\begin{thebibliography}{10}
\providecommand{\url}[1]{\texttt{#1}}
\providecommand{\urlprefix}{URL }

\bibitem{Lifshitz}
I.~Lifshitz,
\newblock \emph{Sov. Phys. JETP} \textbf{1960}, \emph{11}, 5 1130.

\bibitem{vanHove}
L.~Van~Hove,
\newblock \emph{Phys. Rev.} \textbf{1953}, \emph{89} 1189.

\bibitem{Chandrasekaran-Shtyk-Betouras-Chamon}
A.~Chandrasekaran, A.~Shtyk, J.~J. Betouras, C.~Chamon,
\newblock \emph{Phys. Rev. Research} \textbf{2020}, \emph{2} 013355.

\bibitem{Efremov-Betouras}
D.~V. Efremov, A.~Shtyk, A.~W. Rost, C.~Chamon, A.~P. Mackenzie, J.~J.
  Betouras,
\newblock \emph{Phys. Rev. Lett.} \textbf{2019}, \emph{123} 207202.

\bibitem{Shtyk}
A.~Shtyk, G.~Goldstein, C.~Chamon,
\newblock \emph{Phys. Rev. B} \textbf{2017}, \emph{95} 035137.

\bibitem{LiangFu}
N.~F.~Q. Yuan, L.~Fu,
\newblock \emph{Phys. Rev. B} \textbf{2020}, \emph{101} 125120.

\bibitem{Zervou-Goldstein-Efremov-Betouras}
A.~Zervou, G.~Goldstein, D.~V. Efremov, J.~J. Betouras,
\newblock {The fate of density waves in the presence of a higher order van Hove
  singularity}, \textbf{2022},
\newblock \urlprefix\url{https://arxiv.org/abs/2205.08828}.

\bibitem{Aoki}
D.~Aoki, G.~Seyfarth, A.~Pourret, A.~Gourgout, A.~McCollam, J.~A.~N. Bruin,
  Y.~Krupko, I.~Sheikin,
\newblock \emph{Phys. Rev. Lett.} \textbf{2016}, \emph{116} 037202.

\bibitem{Barber}
M.~E. Barber, F.~Lechermann, S.~V. Streltsov, S.~L. Skornyakov, S.~Ghosh, B.~J.
  Ramshaw, N.~Kikugawa, D.~A. Sokolov, A.~P. Mackenzie, C.~W. Hicks, I.~I.
  Mazin,
\newblock \emph{Phys. Rev. B} \textbf{2019}, \emph{100} 245139.

\bibitem{Bernhabib}
S.~Benhabib, A.~Sacuto, M.~Civelli, I.~Paul, M.~Cazayous, Y.~Gallais, M.-A.
  M\'easson, R.~D. Zhong, J.~Schneeloch, G.~D. Gu, D.~Colson, A.~Forget,
\newblock \emph{Phys. Rev. Lett.} \textbf{2015}, \emph{114} 147001.

\bibitem{Khan}
S.~N. Khan, D.~D. Johnson,
\newblock \emph{Phys. Rev. Lett.} \textbf{2014}, \emph{112} 156401.

\bibitem{Coldea}
A.~I. Coldea, S.~F. Blake, S.~Kasahara, A.~A. Haghighirad, M.~D. Watson,
  W.~Knafo, E.~S. Choi, A.~McCollam, P.~Reiss, T.~Yamashita, et~al.,
\newblock \emph{npj Quantum Materials} \textbf{2019}, \emph{4}, 1 1.

\bibitem{Okamoto}
Y.~Okamoto, A.~Nishio, Z.~Hiroi,
\newblock \emph{Phys. Rev. B} \textbf{2010}, \emph{81} 121102.

\bibitem{Sherkunov-Chubukov-Betouras}
Y.~Sherkunov, A.~V. Chubukov, J.~J. Betouras,
\newblock \emph{Phys. Rev. Lett.} \textbf{2018}, \emph{121} 097001.

\bibitem{Slizovskiy-Chubukov-Betouras}
S.~Slizovskiy, A.~V. Chubukov, J.~J. Betouras,
\newblock \emph{Phys. Rev. Lett.} \textbf{2015}, \emph{114} 066403.

\bibitem{Stewart}
I.~Stewart,
\newblock \emph{Reports on Progress in Physics} \textbf{1982}, \emph{45}, 2
  185.

\bibitem{Yelland}
E.~Yelland, J.~Barraclough, W.~Wang, K.~Kamenev, A.~Huxley,
\newblock \emph{Nature physics} \textbf{2011}, \emph{7}, 11 890.

\bibitem{Yuan}
N.~F. Yuan, H.~Isobe, L.~Fu,
\newblock \emph{Nature communications} \textbf{2019}, \emph{10}, 1 1.

\bibitem{Sherkunov-Betouras}
Y.~Sherkunov, J.~J. Betouras,
\newblock \emph{Phys. Rev. B} \textbf{2018}, \emph{98} 205151.

\bibitem{Classen_2020}
L.~Classen, A.~V. Chubukov, C.~Honerkamp, M.~M. Scherer,
\newblock \emph{Phys. Rev. B} \textbf{2020}, \emph{102} 125141.

\bibitem{Isobe}
H.~Isobe, L.~Fu,
\newblock \emph{Phys. Rev. Research} \textbf{2019}, \emph{1} 033206.

\bibitem{doped_graphene}
J.~L. McChesney, A.~Bostwick, T.~Ohta, T.~Seyller, K.~Horn, J.~Gonz\'alez,
  E.~Rotenberg,
\newblock \emph{Phys. Rev. Lett.} \textbf{2010}, \emph{104} 136803.

\bibitem{Rosenzweig_2020}
P.~Rosenzweig, H.~Karakachian, D.~Marchenko, K.~K\"uster, U.~Starke,
\newblock \emph{Phys. Rev. Lett.} \textbf{2020}, \emph{125} 176403.

\bibitem{Kang}
M.~Kang, S.~Fang, J.-K. Kim, B.~R. Ortiz, S.~H. Ryu, J.~Kim, J.~Yoo,
  G.~Sangiovanni, D.~Di~Sante, B.-G. Park, et~al.,
\newblock \emph{Nature Physics} \textbf{2022}, 1--8.

\bibitem{kagome2}
Y.~Hu, X.~Wu, B.~R. Ortiz, S.~Ju, X.~Han, J.~Ma, N.~C. Plumb, M.~Radovic,
  R.~Thomale, S.~D. Wilson, et~al.,
\newblock \emph{Nature Communications} \textbf{2022}, \emph{13}, 1 1.

\bibitem{Consiglio}
A.~Consiglio, T.~Schwemmer, X.~Wu, W.~Hanke, T.~Neupert, R.~Thomale,
  G.~Sangiovanni, D.~Di~Sante,
\newblock \emph{Phys. Rev. B} \textbf{2022}, \emph{105} 165146.

\bibitem{Neupert}
T.~Neupert, M.~M. Denner, J.-X. Yin, R.~Thomale, M.~Z. Hasan,
\newblock \emph{Nature Physics} \textbf{2022}, \emph{18}, 2 137.

\bibitem{Zhou_2021}
H.~Zhou, L.~Holleis, Y.~Saito, L.~Cohen, W.~Huynh, C.~L. Patterson, F.~Yang,
  T.~Taniguchi, K.~Watanabe, A.~F. Young,
\newblock \emph{Science} \textbf{2022}, \emph{375}, 6582 774.

\bibitem{Chandrasekaran-Betouras}
A.~Chandrasekaran, J.~J. Betouras,
\newblock \emph{Phys. Rev. B} \textbf{2022}, \emph{105} 075144.

\bibitem{Kokkinis}
E.~K. Kokkinis, G.~Goldstein, D.~V. Efremov, J.~J. Betouras,
\newblock \emph{Phys. Rev. B} \textbf{2022}, \emph{105} 155123.

\bibitem{abel}
R.~G. Ayoub,
\newblock \emph{Archive for history of exact sciences} \textbf{1980}, 253--277.

\bibitem{bradlyn2016beyond}
B.~Bradlyn, J.~Cano, Z.~Wang, M.~Vergniory, C.~Felser, R.~J. Cava, B.~A.
  Bernevig,
\newblock \emph{Science} \textbf{2016}, \emph{353}, 6299 aaf5037.

\bibitem{Mathematica}
W.~R. Inc.,
\newblock {Mathematica, Version 13.1},
\newblock \urlprefix\url{https://www.wolfram.com/mathematica},
\newblock Champaign, IL, 2022.

\bibitem{Maple}
{Maplesoft, a division of Waterloo Maple Inc..},
\newblock Maple,
\newblock \urlprefix\url{https://hadoop.apache.org}.

\bibitem{HaldaneModel}
F.~D.~M. Haldane,
\newblock \emph{Phys. Rev. Lett.} \textbf{1988}, \emph{61} 2015.

\bibitem{bandutils}
A.~Chandrasekaran,
\newblock Band utilities,
\newblock
  \url{https://github.com/anirudhc-git/Series-expansion-of-bandstructure},
\newblock A Mathematica package for analysing bandstructures arising from
  k-space models.

\end{thebibliography}

\end{document}